\documentclass{aa}
\pdfoutput=1
 
\usepackage{amsmath}
\usepackage[varg]{txfonts}

\usepackage{longtable} 
\usepackage{pdflscape} 
\usepackage{graphicx}
\usepackage{natbib}
\usepackage{xspace}
\usepackage{nicefrac}
\usepackage{aas_macros}
\usepackage{hyperref}  
\usepackage{multirow}
\usepackage[usenames, dvipsnames]{color}  
\usepackage{tablefootnote}
\defcitealias{Podsiadlowski+1992}{P92}
\defcitealias{Zapartas+2017}{Z17}
\defcitealias{Zapartas+2019}{Z19}

\newcommand{\Msun}{\ensuremath{M_{\odot}}\xspace}

\authorrunning{Zapartas et al.}
\titlerunning {Mass distribution of binary progenitors of type II SNe}

\begin{document}

\title{Effect of binary evolution on the inferred initial and final core masses of hydrogen-rich, Type~II supernova progenitors}

\author{E. Zapartas\inst{1,2,*}, S.\,E. de Mink\inst{3,2}, S. Justham\inst{4,5,2}, N. Smith\inst{6}, M. Renzo\inst{7,2} \& A. de Koter\inst{2,8}} 

\institute{Geneva Observatory, University of Geneva, Chemin des Maillettes 51, 1290 Sauverny, Switzerland.
\and Anton Pannenkoek Institute for Astronomy, University of Amsterdam, 1090 GE Amsterdam, The Netherlands
  \and Center for Astrophysics, Harvard-Smithsonian, 60 Garden Street, Cambridge, MA 02138, USA
  \and  School of Astronomy, Space Science, University of the Chinese Academy of Sciences, Beijing 100012, China.
  \and National Astronomical Observatories, Chinese Academy of Sciences, Beijing 100012, China.
 \and Steward Observatory, University of Arizona, 933 N. Cherry Avenue, Tucson, AZ 85721, USA.
  \and Center for Computational Astrophysics, Flatiron Institute, New York, NY 10010, USA
 \and Institute of Astronomy, KU Leuven, Celestijnenlaan 200 D, B-3001 Leuven, Belgium.
    \\ $^*$\email{ezapartas@gmail.com}\\
  }

\abstract{
The majority of massive stars, which are the progenitors of core-collapse supernovae (SNe), are found in close binary systems. In a previous work, we modeled the fraction of hydrogen-rich, Type II SN progenitors whose evolution is affected by mass exchange with their companion, finding this to be between $\approx$ 1/3 and 1/2 for most assumptions. Here we study in more depth the impact of this binary history of Type II SN progenitors on their final pre-SN core mass distribution, using population synthesis simulations. 
We find that binary star progenitors of Type II SNe typically end their life with a larger core mass than they would have had if they had lived in isolation because they gained mass or merged with a companion before their explosion. 
The combination of the diverse binary evolutionary paths typically leads to a marginally shallower final core mass distribution. In discussing our results in the context of  the red supergiant problem, that is, the reported lack of detected high luminosity progenitors, we conclude that binary evolution does not seem to significantly affect the issue. This conclusion is quite robust against our variations in the assumptions of binary physics. 
We also predict that inferring the initial masses of Type II SN progenitors by ``age-dating'' their surrounding environment systematically yields lower masses compared to methods that probe the pre-SN core mass or luminosity. A robust discrepancy  between the inferred initial masses of a SN progenitor from those different techniques could indicate an evolutionary history of binary mass accretion or merging. 
}


\date{    Published in Astronomy \& Astrophysics }

\maketitle

\section{Introduction}\label{ch6:sec:intro}

Core-collapse supernovae (SNe) occur at the end of the evolution of massive stars \citep[e.g.,][]{Baade+1934,Bethe+1979,Woosley+2002,Heger+2003}.  They are essential for the chemical enrichment of galaxies and as sources of feedback \citep[e.g.,][]{Larson1974,Hopkins+2014}. Supernovae are classified observationally into two main groups \citep[e.g.,][]{Filippenko1997, Gal-Yam2017}: hydrogen-rich, Type II SNe (including types II-P, II-L and IIn) and stripped-envelope SNe, which either show the complete absence of hydrogen features in their spectra or which show evidence for hydrogen only at early times (including types IIb, Ib, Ic, and more exotic examples such as Type Ic-broadline and Ibn). The first group are thought to  arise from massive stars which have retained most of their hydrogen-rich envelope until the explosion.  This study focuses on the progenitors of these hydrogen-rich, Type II SNe (hereafter referred to as SN~II). 

The reason why some massive stars end their lives with most of their hydrogen-rich envelope still intact and others do not is not fully understood. The canonical picture considers the progenitor stars as single stars that evolve in isolation. They can lose their envelope through stellar-wind mass loss or eruptive mass loss episodes \citep{Conti1975}. In this picture, SN~II are proposed to originate from single stars with initial masses between $\sim 8$ and $\sim 25-30$ \Msun  at solar metallicity  \citep[e.g.,][]{Heger+2003, Eldridge+2004, Georgy+2012, Renzo+2017}. 

Massive stars with initial masses in this range are very often found in close binary systems \citep{Kobulnicky+2007, Eggleton+2008, Chini+2012, Kiminki+2012, Sana+2012,  Dunstall+2015, Moe+2017, Almeida+2017}.  A large fraction of them are expected to interact with their stellar companion during their lifetime.  Such interactions can remove the envelope of the donor star and give rise to a stripped-envelope SNe. This has been the topic of various theoretical studies \citep[e.g.,][and references therein]{Podsiadlowski+1992, De-Donder+1998, Yoon+2010, Claeys+2011, Eldridge+2013, Yoon+2017, Zapartas+2017a, Sravan+2018,Laplace+2020} 
in order to help explain their high relative rates \citep[e.g.,][]{Smartt+2009, Smith+2011, Li+2011, Eldridge+2013, Graur+2017}, their low ejecta masses \citep[e.g.,][]{Drout+2011, Taddia+2015, Lyman+2016,Prentice+2019}, and the difficulty of finding their progenitors \citep[e.g.,][]{Eldridge+2013, Cao+2013, Van-Dyk+2018}.

The possible role of binary interaction in the lives of hydrogen-rich SN~II is less explored, but this topic is gaining interest  \citep[e.g.,][]{Podsiadlowski+1992,Podsiadlowski1992, Tutukov+1992, Vanbeveren+2013, Justham+2014, Menon+2017, Menon+2019, Eldridge+2018, Eldridge+2019}.  In a previous study \citep[][hereafter referred to as \citetalias{Zapartas+2019}]{Zapartas+2019}, we used population synthesis simulations to investigate the rates of different binary evolutionary channels that can lead toward hydrogen-rich Type II SN progenitors.  Our simulations presented in that work suggest that a significant fraction of all SN~II progenitors, $\sim$30-50\%, are expected to have interacted with their companion prior to explosion. This prediction is robust against several variations in the model assumptions. Almost all of these binary progenitors are either stars that gained mass through Roche-lobe overflow (RLOF) or stars that merged with their companion.  The remainder of the SN~II progenitors evolve without experiencing any significant type of interaction. They effectively evolve as single stars in isolation.  These may be either true single stars or stars that have one or more companions orbiting too far away for mass transfer to occur.

If binary interaction indeed plays a significant role in the lives of the population of SN~II progenitors, as our previous simulations suggest, then this would have various implications that deserve further investigation.   For example, we would expect a diverse set of observable consequences, such as possible influence on their light curves and systematic differences in the offset between the location of the explosions and the nearest star-forming region (see \citealt{Eldridge+2018}, \citetalias{Zapartas+2019} for a discussion).   

The aim of this work is to investigate, in a statistical manner, the expected impact of binary evolutionary channels on SN~II progenitors. We focus on the distribution of final core masses. 
The final core mass is one of the primary parameters that characterize the final state of a massive star. It is thought to play a key role in determining its fate and is a primary parameter that affects the late pre-SN observables such as the luminosity of the progenitor \citep[e.g.,][]{Giannone1967,Sukhbold+2016,Farrell+2020}.   Other parameters, such as the details of the internal density, composition and angular momentum profile and magnetic field strength also play a role, but at present it is much harder, or even impossible, to make reliable statistical predictions for these quantities. Even the prediction of final core masses for stars that are the products of binary interaction is not free from caveats, but we consider this as a natural starting point.  

We first present the initial - final core mass relation that holds for single stars and show how binarity breaks this one-to-one relation.  We find that Type II SN progenitors in binary systems typically originate from lower initial masses than single stars that reach similar core masses, because they gained mass or merged with a companion before explosion. 
We find that binary star progenitors of SN~II typically end their life with a larger core mass than they would have had if they had lived in isolation.   We investigate the contribution of different single and binary evolutionary channels and show the implications for the distribution of final core masses  expected for  Type II SNe.  

We then place our theoretical results in context of the observations, using the sample of SN~II compiled by \citet{Smartt+2009,Smartt2015}. 
This sample consists of 26 nearby SN~II, for which high resolution, archival images are available of the explosion site. These images have been used to derive  constraints on the properties of the progenitors  \citep[see also][]{Van-Dyk+2003, Maund+2005a}. We compare our predictions with these constraints, discuss the validity and limitations of this comparison and the impact of our model assumptions.  We further discuss how binarity affects the so-called red supergiant problem, that is, the claimed  lack of red supergiant progenitors with initial masses above $\sim 17 \, $\Msun \citep[][but see also \citealt{Davies+2018,Davies+2020}]{Smartt+2009}.  
We also discuss the implications of binary interaction in other techniques that infer the initial mass of Type II progenitors and predict a systematic discrepancy among them.

The paper is structured as followed. We explain our method in section~\ref{ch6:sec:model}, present our results in section~\ref{ch6:sec:results} and place them in context of SN progenitor detections (section~\ref{ch6:sec:rsg_problem}), after exploring the sensitivity of our findings to model assumptions. We discuss the implications of our results for other observational constraints of Type II SN progenitors in section~\ref{ch6:sec:other_observational_constraints} and summarize our conclusions in section \ref{ch6:sec:summary}. Our paper can be considered as a companion study to \citetalias{Zapartas+2019}, but can be read independently.


\section{Method}\label{ch6:sec:model}

In order to study the end fate of Type II SN progenitors we simulate the evolution of entire populations of single and binary stellar systems. 
For these simulations we employ the stellar population synthesis code {\tt binary\_c}, developed by \citet{Izzard+2004,Izzard+2006,Izzard+2009} with updates described in \citet{de-Mink+2013} and \citet{Schneider+2015}. In this code, the evolution of the stars is based on analytical fitting formulae \citep{Hurley+2000} derived from the grid of detailed simulations of  \citet{Pols+1998}. \citet{Hurley+2002} presented the implementation of  these evolutionary formulae in the context of population  synthesis simulations, which consider also physical processes in binary systems. 

This code allows us to explore the initial parameter space of our population, which is very extended due to the numerous possible configurations of binary systems. Using this code, we have created a grid of simulations to study the statistical properties of core-collapse SNe \citep[][hereafter \citetalias{Zapartas+2017}]{Zapartas+2017}. In this study we focus on models describing the evolution of Type II SN progenitors that retained their hydrogen-rich envelope until the explosion, as in the case of \citetalias{Zapartas+2019}. 
Our main grid consists of $10^4$ single stars of various initial masses and $150\times 150 \times 150$ binary systems varying their initial mass, their mass ratio and their orbital period.
We assign a probability of formation to each of these stellar systems according to our assumed initial distributions. For our standard assumptions we follow an initial mass function (IMF) for our single and primary stars from \citet{Kroupa2001}, and an initial mass ratio and period distribution following \citet{Sana+2012} for initially O-type primaries. The latter becomes flatter for initially lower mass systems \citep{Opik1924}. The slope of the initial distributions above are referred to as $\alpha_{\rm IMF}$, $\kappa$, and $\pi$, respectively. Our results are described in the context of populations that consist exclusively of single stars or alternatively of binary systems, but we present also our findings for more realistic populations that include a combination of these extreme cases, assuming a binary fraction of $f_{\rm bin} = 0.3, 0.5, 0.7$ and a mass dependent case $f_{\rm bin}(M)$, as in \citetalias{Zapartas+2017}.

The speed of population synthesis codes allows us to investigate the sensitivity of our main findings to variations in our input assumptions, testing the robustness of our results. These assumptions concern the initial stellar population, as specified above, as well as the treatment of the physical processes that take place during the evolution of single and binary stars. We first present the results of our fiducial model, in which we choose a standard set of assumptions that we describe here.  
For a more detailed description of all the input model assumptions we refer to \citetalias{Zapartas+2017}. 
We also perform simulations in which we vary our main model assumptions one-by-one, as in \citetalias{Zapartas+2017}, \citetalias{Zapartas+2019}. An overview is given in Table~\ref{ch6:table:parameters_uncertainties}.  For consistency, we link and refer to the model variations  that are also presented in \citetalias{Zapartas+2017} with the same model number.

To account for the effect of mass loss on the evolution of stars, we follow the standard rate prescriptions from \citet{Vink+2000,Vink+2001} and \citet{Nieuwenhuijzen+1990} during different evolution phases of hydrogen-rich stars, as well as the rate of \citet{Hamann+1995} reduced by a factor of 10 for stripped helium cores \citep{Yoon+2005}. We also consider the effect of varying wind mass loss efficiency, $\eta$, between the range of 0.1 to 3 (Models 25, 26 and 51-54). The chosen metallicity for our main results is solar-like \citep[$Z=0.014$]{Asplund+2009} although we also consider different, mainly lower metallicities (Models 40, 42, 44).  

We treat stellar spin following \citet{Hurley+2000} and account for the effects of tides in binary systems \citep{Zahn1977, Hut1980, Hut1981, Hurley+2002}. We simulate stable Roche-lobe overflow removing mass from the donor so as to remain inside its Roche lobe but limiting its mass loss rate according to its thermal rate. The accretion of matter from the companion is limited to 10 times its thermal rate \citep{Schneider+2015} although we run also simulations in which we vary the efficiency of this process as a fixed parameter ($\beta$, Models 1 to 3). The mass lost from the system is assumed to carry the specific orbital angular momentum of the accretor, with variations also ranging from no angular momentum loss to losses as high as to form a Keplerian circumbinary disk with a radius described in \citetalias{Zapartas+2017} (parameter $\gamma$, Models 4 and 5). 
We account for rejuvenation of the accreting star or merger according to \citet{Tout+1997,de-Mink+2013} and \citet{Schneider+2015}, considering variations of the fraction of mass lost and mixed in case of main-sequence mergers (parameters $\mu_{\rm loss}$ and $\mu_{\rm mix}$ respectively, Models 6 to 9).

We consider the possibility of unstable mass transfer in cases of mass ratios more extreme than some critical value ($M_{\rm accretor}/M_{\rm donor} <q_{\rm crit}$), depending on the evolutionary state of the donor. This accounts for the possibility of a runaway mass transfer process as well as the case of simultaneously filling of the Roche lobes by both stars. We follow \citet{Hurley+2002} for the mass-ratio threshold of $q_{\rm crit}$, apart from the case of a donor crossing the Hertzsprung gap where we follow \citet{de-Mink+2013}. In the case of two main sequence stars, such an instability leads to contact and merging, while in the other cases it leads to common envelope evolution. We consider different values of the mass ratio stability criteria  (see Table~\ref{ch6:table:parameters_uncertainties}). We focus mostly on variations during the early evolutionary phases of the donor and have not explored the uncertainty of stability for Red Supergiant donors in initially wide orbits \citep[e.g.,][]{Ge+2010,Shao+2014,Pavlovskii+2015,Pavlovskii+2017} as the number of  type II progenitors originating from the latter is low in our fiducial simulation \citepalias[Fig. 4 in][]{Zapartas+2019}. 

The outcome of the CE is the spiral-in of the components resulting either in a merger or the ejection of the common envelope, according to the $\alpha_{CE}$-prescription \citep{Webbink1984}. In our fiducial model we assume $\alpha_{CE} =1 $ and a gravitational binding energy parameter $\lambda_{CE}$ according to \citet{Dewi+2000,Dewi+2001} and \citet{Tauris+2001}. We consider other values for both in our variations (Table~\ref{ch6:table:parameters_uncertainties}).

We account for supernova kicks, both due to instantaneous mass loss \citep{Blaauw1961} as well as a possilbe natal kick to the remnant due to asymmetries in the explosion. These kicks may disrupt a  binary system in most of the cases \citep{Eldridge+2011,Renzo+2019}. In our fiducial model we select random natal kicks from a Maxwellian distribution of root-mean square $\sigma = 265 \rm{km \, s^{-1}}$, based on the work of \citep{Hobbs+2005} on observed radio pulars. Natal SN kicks are uncertain, with  evidence for lower kicks in some cases \citep[e.g.,][]{Verbunt+2017}. To test the sensitivity of our results to this assumption we also consider extreme variations of no and or very high natal kicks, affecting the rate of disrupted binary systems and thus of mass gainer type II SN progenitors \citepalias{Zapartas+2019}. For a more in depth study of the impact of kick to star ejected from binary systems, we refer the reader to \citet{Renzo+2019}.  If the system remains bound, we recalculate its orbital parameters. In our model variations we also consider the possibility of core-collapse events resulting in direct fallback onto a black hole with no observable transient \citep[e.g.,][]{OConnor+2011,Sukhbold+2016,Adams+2017} for cases in which the final core of the progenitor is more massive than the core mass corresponding to a single star of $M_{\rm max,ccSN}$ (Models 29, 52, 53, 54). We also run simulations where we artificially change the core mass criterion for a collapse, essentially slightly varying the minimum mass threshold for a core-collapse SN, $M_{\rm min,ccSN}$ (Models 30 and 31).

\begin{figure*}[t]\center
\includegraphics[width=0.50\textwidth]{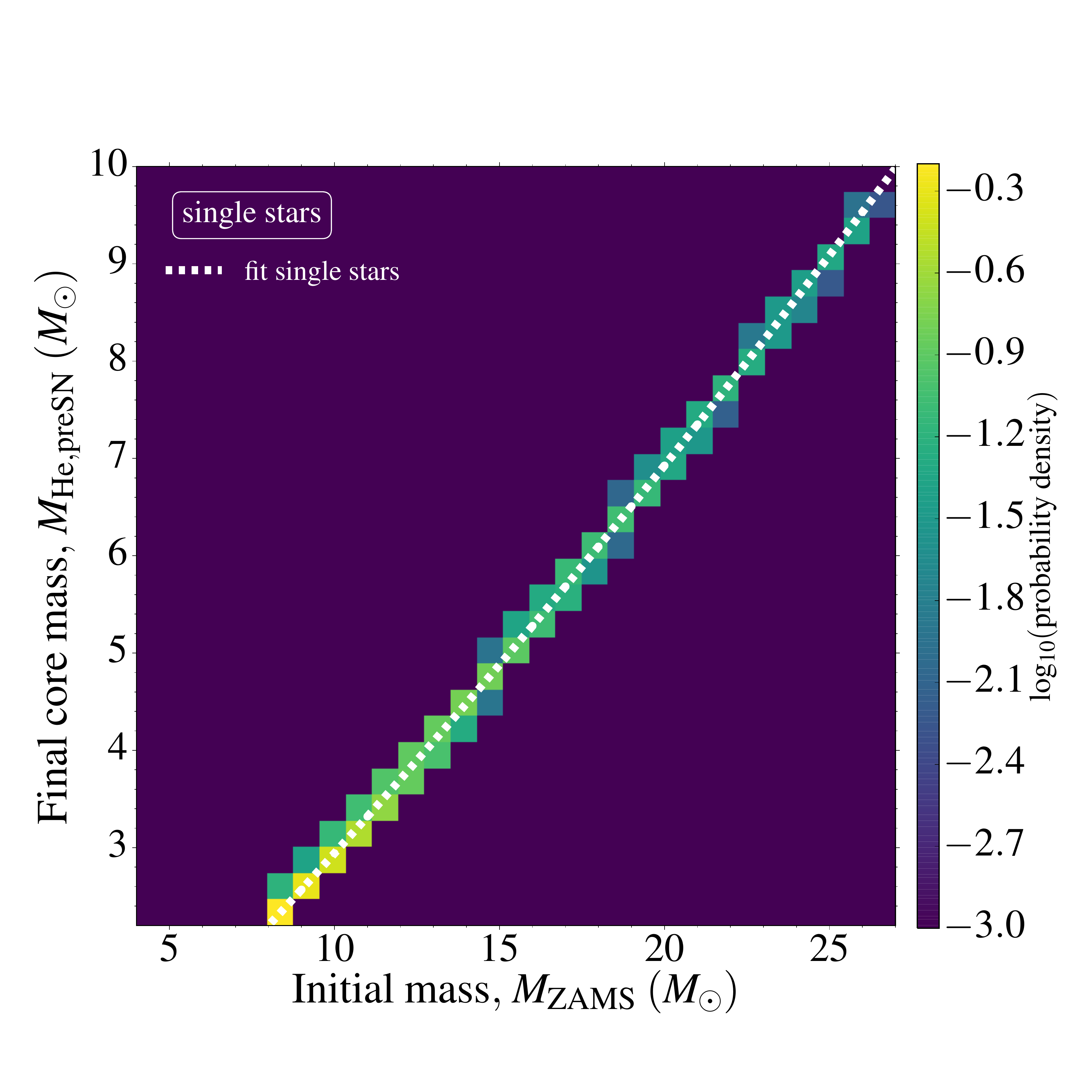}\includegraphics[width=0.50\textwidth]{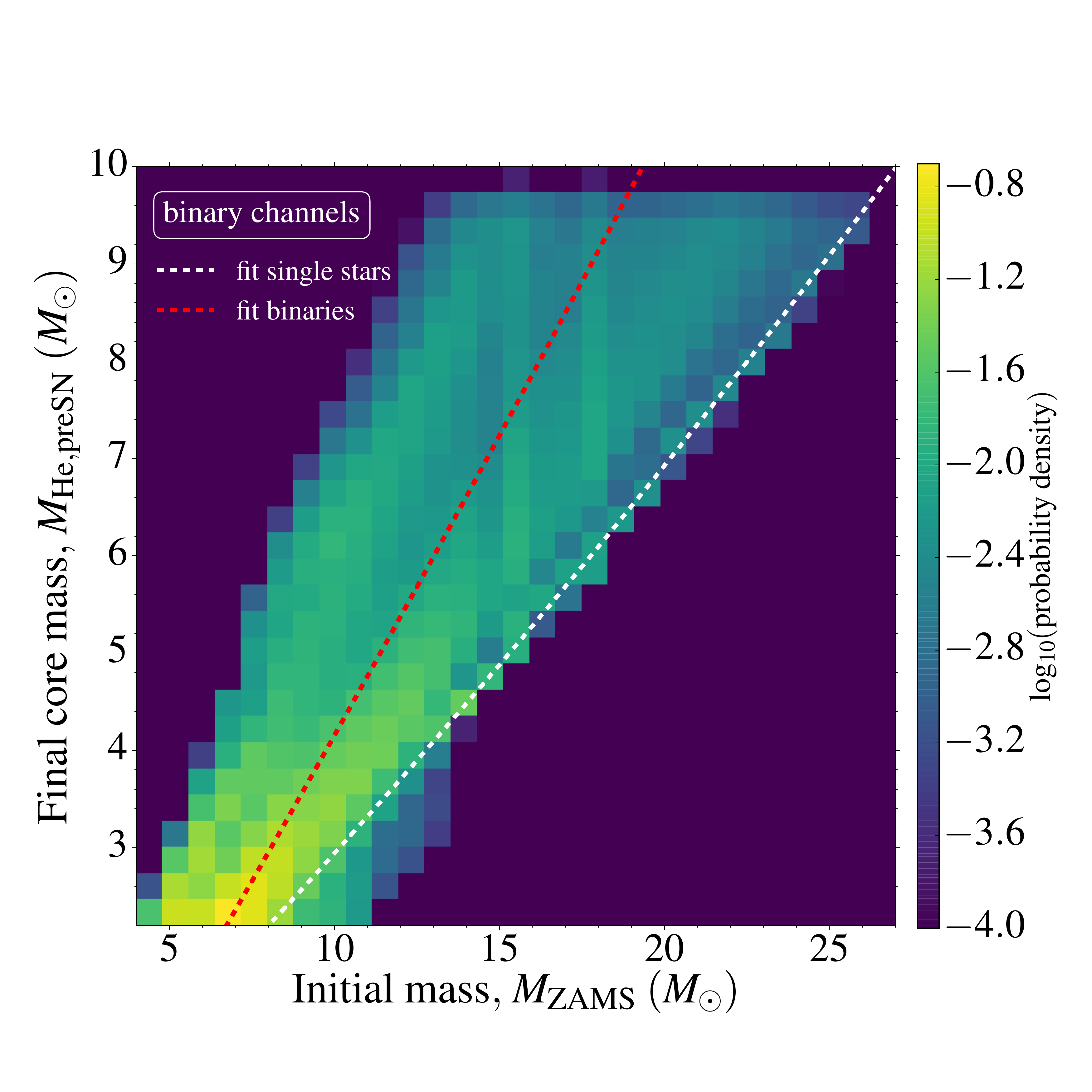}
\caption{
These diagrams show how binary interaction broadens the initial to final core mass relation of SN~II progenitors. We present the normalized 2D density distribution of the final core mass, $M_{\rm He,preSN} $, with the initial mass, $M_{\rm ZAMS}$, for all SN~II progenitors in case they have an evolution in isolation (left panel) or in case they experience mass exchange with a binary companion (right panel). The color mapping is in logarithmic scale. We also depict the fitting formula for the one-to-one relation of single stars (Eq.~\ref{ch6:eq:helium_zams_singles}; white dashed line) and for the median of the initial mass distribution for the case of binary progenitors (Eq.~\ref{ch6:eq:helium_zams_binaries}; red dashed line).
}
\label{ch6:fig:2D_hecore_ZAMS}
\end{figure*}

\section{Results}\label{ch6:sec:results}

In this section we discuss the impact of binary interaction on the distribution of the final core masses of Type II SNe progenitors.   Throughout this section and those that follow, we use the term ``final core mass'' to refer to the final mass of the helium core or, more accurately, the hydrogen-depleted core,  $M_{\rm He,preSN}$, of the supernova progenitor immediately prior to the explosion as predicted in our simulations.  We use the term ``initial mass'' to indicate the zero-age main sequence mass in the case of single stars, $M_{\rm ZAMS}$.  In the case of binary systems, we define $M_{\rm ZAMS}$ as the mass of the initially more massive primary star, unless the SN originates from the initially less massive secondary star.  

We first discuss the relation between the final core mass and the initial mass for single stars and how binaries affect this (section \ref{ch6:sec:results_helium_zams_relation}). We then show the impact of different binary evolutionary channels (section~\ref{ch6:sec:results_role_channels}). We end with our predictions for the distribution of final core masses as we would expect from an idealized survey of observed Type II SNe events (section~\ref{ch6:sec:results_slope}).

\subsection{Binary progenitors of Type II SNe  typically originate from lower mass stars than single star progenitors }
\label{ch6:sec:results_helium_zams_relation}

For single stars there is a tight relation between the initial mass of a star and its final core mass.  This is because the evolution and final fate of a single star is primarily a function of its initial mass.  Other parameters, such as for example, the metallicity and the initial rotation rate play a role, but are less important \cite[e.g.,][]{Heger+2003, Brott+2011a}. This statement breaks when considering extreme values, but such cases are very rare, certainly for events in the local universe. This means that for a given set of assumptions (for various uncertain internal physical processes, such as convective overshooting, rotational mixing, etc), the initial-final core mass relation is effectively a one-to-one mapping.

The left panel of Figure~\ref{ch6:fig:2D_hecore_ZAMS} shows the initial - final core mass relation for our fiducial simulation of single-star progenitors of Type II SNe.  Under the assumptions followed in these simulations, Type II SNe result from stars with initial masses in the range  between $M_{\rm min,ccSN} \approx 7.6$  \Msun and $M_{\rm min,WR}   \approx 27$ \Msun.    Stars that are initially less massive than  $M_{\rm min,ccSN}$ end their life as white dwarfs, whereas the ones more massive than $M_{\rm min,WR}$ lose their hydrogen-rich envelope due to winds prior to explosion and become Wolf-Rayet stars. If they explode successfully, they would result in stripped-envelope SNe instead of the Type II SNe that we focus on here.   The shape of the relation and the mass boundaries depend on the physical assumptions, in particular the treatment of internal mixing processes and stellar wind mass loss. However, for fixed assumptions, we always expect this relation between the initial mass and the final core mass to be tight for single-star progenitors. 

Mass exchange in binary systems can radically affect the pre-SN structure of a massive star \citep[e.g.,][]{Podsiadlowski+1992} and therefore breaks the one-to-one mapping between the initial stellar mass and the final core mass of a SN progenitor. This is illustrated in the right panel of Figure \ref{ch6:fig:2D_hecore_ZAMS}, which shows a very broad distribution for binary progenitors.   

We find that binary star progenitors of Type II SNe typically end their life with a larger core mass than they would have had if they had lived in isolation. This can be seen more clearly by comparing the distribution for binary progenitors with the white dashed line, which shows an analytical fit to the relation we obtained for single stars. This can occur, for example, after gaining mass from their companion through stable mass transfer or after  a  merger (as we show in more depth in section~\ref{ch6:sec:results_role_channels}). If such mass gain happens early in the evolution of the progenitor, we expect that its core can still grow as the star adapts its interior structure to its new mass \citep{Hellings1983,Hellings1984,Braun+1995, Dray+2007}.  The star, therefore, ends its life with a core mass that is larger  than if it would have been a single star. 

The red dashed line shows a similar analytical fit to the median of the initial mass distribution for binary progenitors. The offset between the white and red dashed line emphasizes the systematic trend that binary progenitors tend to end their lives with larger core masses than expected for single stars with the same initial mass.  Stating this more quantitatively: for a given final core mass, we find that binary progenitors of Type II SNe typically start their lives with an initial mass that is 15-25\% lower than single stars.  This implies that if one would not account for the possibility that a Type II SN progenitor experienced binary interaction, then one would typically overestimate its true initial mass. 

The fact that almost no progenitors end their lives with core masses that are lower than expected according to the single star relation is a consequence of several effects.  First of all, the mass of the core of a star is largely determined   early in its evolution, when the star has completed its main sequence phase.  
After this a typical progenitor from a binary route seems unlikely to significantly reduce the mass of its core compared to a single star progenitor. Although stars may be spun-up by binary interactions, the transition between helium core and hydrogen-rich envelope is marked by a sharp chemical gradient, which largely inhibits rotational mixing that might otherwise reduce the mass of the core \citep[e.g.,][]{Mestel1957,Mestel+1986}. However, in some cases other mixing processes can be effective in eroding the core of a merged star \citep[e.g.,][]{Justham+2014}, which we neglect in this study. 
In principle, mass loss from the star could be also thought of as a mechanism to reduce the core size, but mass loss primarily affects the decoupled envelope. The core is not significantly affected unless the envelope is removed almost entirely and in this study we focus on Type II SNe, which still have at least part of their hydrogen envelopes left until the moment of explosion.  
Only for progenitors with low final core masses, with $M_{\rm He,preSN} \lesssim 4\Msun$, we find binary progenitors to lie below the single star relation  as can be seen in Figure \ref{ch6:fig:2D_hecore_ZAMS}.  Most of these systems originate from mergers of two evolved stars, called reverse mergers, that we briefly discuss in the next subsection. 

For convenience we provide the analytical fit to the one-to-one relation between the initial mass, $M_{\rm ZAMS}$, and the final core mass, $M_{\rm He,preSN}$ for single stars, depicted by the white dashed lines in Figure \ref{ch6:fig:2D_hecore_ZAMS}. 
We find that the single-star relation is well described  (with a relative error of $<0.5\%$) by the second-order polynomial,
\begin{equation}\label{ch6:eq:helium_zams_singles}
\frac{M_{\rm He,preSN, singles}}{\Msun} = 0.33 \frac{M_{\rm ZAMS}}{\Msun}+ 0.002 (\frac{M_{\rm ZAMS}}{\Msun})^2 - 0.59, 
\end{equation}
in the range of $8<M_{\rm ZAMS}/\Msun <25$. 
Later in this study, we use the inverse of Equation~\ref{ch6:eq:helium_zams_singles} to attribute to each pre-SN core mass an ``equivalent single-star initial mass'' that would have formed such a massive core if it was in isolation, even in cases that the progenitor in reality exchanged mass with its companion.

For binary progenitors the distribution is very broad, but if we want to describe the median of the distribution with a similar polynomial, we get:
\begin{equation}\label{ch6:eq:helium_zams_binaries}
\frac{M_{\rm He,preSN, binaries}}{\Msun} = 0.56 \frac{M_{\rm ZAMS}}{\Msun}+ 0.002 (\frac{M_{\rm ZAMS}}{\Msun})^2 - 1.67, 
\end{equation}
in the range of $7<M_{\rm ZAMS}/\Msun<18$, with a relative error of  $<9\%$ between the fitting formula and the median of the distribution. 
Eq.~\ref{ch6:eq:helium_zams_binaries} is shown as the red dashed line in the right panel of Figure \ref{ch6:fig:2D_hecore_ZAMS}.

\begin{figure*}[t]\center
\includegraphics[width=0.99\textwidth]{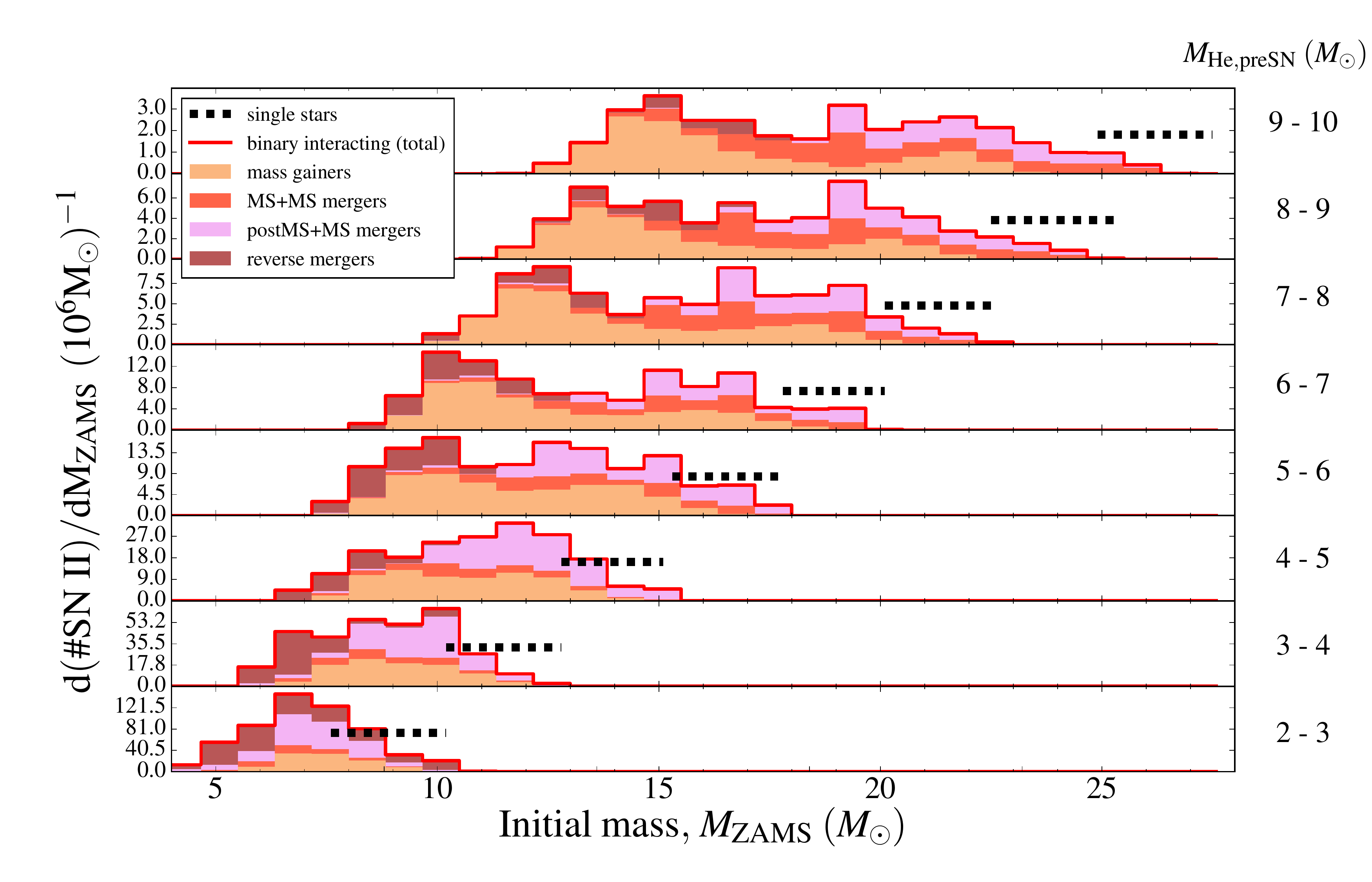}
\caption{Distribution of initial masses, $M_{\rm ZAMS}$, of SN~II progenitors experiencing  binary interaction (red line) for increasing ranges of final  core masses, $M_{\rm He,preSN} $ (on the right part). 
The contribution of each of the main binary scenarios is also shown in color below the distribution of binary-interacted progenitors. 
For reference we include the initial mass range of a single star corresponding to each $M_{\rm He,preSN}$ bin (black dashed line), which is shown in an arbitrary height.  
It becomes clear that the binary progenitors of SN~II can originate from a wider range of lower  on average $M_{\rm ZAMS}$  compared to single stars to form cores of similar mass, due to mass accretion or merging during their evolution. The results are for our standard assumptions although similar trends are found in all our model variations. The values on the y-axis show the number of SN~II per bin in $M_{\rm ZAMS}$ per $10^6$\Msun of stars formed. 
}
\label{ch6:fig:ZAMS_mass_forbinnedhecoremass_typeII}
\end{figure*}

\subsection{Role of different binary channels}\label{ch6:sec:results_role_channels}

The progenitors of Type II SNe originate from a variety of binary-evolution channels, which include stars that gain mass through Roche-lobe overflow and also various types of mergers.  We presented an extensive discussion of these channels in  \citetalias{Zapartas+2019}.  Here, we briefly summarize and discuss how they contribute to the spread in the relation between the initial and final core mass.

To discuss the trends with mass, we choose to subdivide the progenitors into groups that each have similar final core masses.   For each group we show the distribution of the initial masses of their binary progenitors.  The group with the lowest final core masses, between $2-3\Msun$, is shown in the bottom panel of Figure~\ref{ch6:fig:ZAMS_mass_forbinnedhecoremass_typeII}. The panels above show the results for progenitors that end their life with higher final core masses, increasing in steps of 1$\Msun$, up to progenitors with final core masses between $9-10\Msun$, which are shown in the top panel.   The contributions of the different dominant binary channels are shown in different colors.  The range of initial masses of single stars that would have reached the same core mass range is marked with a black dashed line. Its height has no physical meaning.

The figure highlights once more that  binary progenitors with very similar final core masses originate from a wide range of initial masses.  The main diagonal trend that is visible across all diagrams is caused by the fact that progenitors with higher final core masses (shown in the top panels) originate on average from initially more massive stars, both in the case of single stars and stars in binary systems. It is also important to note the decrease of the total number of SNe for higher core masses, mostly due to the initial mass function. The width of the distribution increases for systems with higher final core masses. This is because they typically result from more massive binary systems, where the amount of mass that a progenitor can gain is larger. The figure also shows that progenitors with very similar final core masses originate from very different evolutionary histories (see legend of Figure~\ref{ch6:fig:ZAMS_mass_forbinnedhecoremass_typeII}). 

An important binary channel that leads to Type II SN progenitors \citepalias{Zapartas+2019} consists of secondary stars that have gained mass through stable mass transfer (marked as mass gainers, shown in dark yellow). These are systems in which the binary is disrupted due to the prior explosion of the primary. This leaves the secondary as a single ``walkaway'' or ``runaway'' star \citep[e.g.,][]{Renzo+2019}. Since it is not longer bound to a binary companion, it has the space to swell to giant dimensions at the end of its life without initiating a new phase of reverse mass transfer.  This channel is important for progenitors that end their lives with high final core masses, as can be seen in the top panels.   Typically they originate from stars that are initially half as massive as single stars.  The contribution of this channel is smaller for progenitors with lower core masses, as can be seen in the bottom panels.  This is because the accretor needs to become massive enough to explode as Type II SN  and also due to the requirement that the primary star needs to be massive enough to explode in the first place and unbind the binary system.  We also see in the bottom panels that the mass gainers of lower mass originate from a narrow initial mass range. This is because there is less mass available to accrete in these systems. 

We also find a significant contribution of progenitors that result from channels that involve the merger of two stars.  We distinguish whether it concerns the merger of two main-sequence stars (listed as MS+MS mergers, shown in orange), a post-main-sequence star with a main-sequence one  (listed as postMS+MS mergers, shown in violet) or mergers where both stars are evolved and the merger is initiated by the evolutionary expansion of the secondary star (listed as reverse mergers, shown in brown).  We caution that the evolution and final fate of merger products is particularly uncertain. For a discussion and overview of the limitations of the model predictions we refer to \citetalias{Zapartas+2017} and \citetalias{Zapartas+2019}.

Mergers of the first type, MS+MS, occur early in the evolution of both stars. The resulting merger product is a star whose evolution may resemble that of a more massive single star \citep{Schneider+2019}.  It ends its life with a final core that is more massive than the individual stars would have had if they had evolved in isolation.  These mergers play a more important role in the production of SN progenitors with more massive final cores, as can be seen in the top panels. This is because more massive stars are more often found in very close binary systems, where the stars start to interact already in the main sequence \citep[e.g.,][]{Sana+2012} and because massive stars expand more than lower mass ones during their main sequence, triggering the RLOF and the eventual merging.  

Type II SN progenitors that are products of mergers of an evolved star with its main sequence companion (postMS+MS) originate from binary systems with the smallest difference in initial mass to single stars, compared to the other binary channels. The evolved primary star in this path has already formed a distinct helium core at the moment of merging and so we assume that the steep chemical gradient at the core-envelope boundary suppresses rotational mixing across the boundary \citep{Mestel1957,Mestel+1986}.  Further, the rate at which a post-merger helium core grows in mass in our models is mainly determined by the helium core mass itself.  So, in our simulations, these progenitors lead to a pre-SN helium core mass similar to that of a single star only a few solar masses more massive than that of the initial primary-star mass.   However, this neglects the possibility that the helium core mass may be reduced as a consequence of the merger, for example, by convective dredge-up during the post-merger relaxation phase \citep{Justham+2014}. Indeed, models of the blue supergiant progenitor of the Type II-peculiar SN 1987A, which is thought to have been produced by this type of binary merger \citep{Podsiadlowski+1990, Podsiadlowski1992}, need to invoke extra mixing and so reduction of the core mass \citep{Podsiadlowski1992,Menon+2017}. However, the core erosion found in the \citet{Justham+2014} merger models is less significant for lower amounts of mass accretion during the merger process, and these are the mergers that eventually lead to red supergiant progenitors such as the ones in the observational sample to which we compare our models in section~\ref{ch6:sec:rsg_problem}. 

Mergers of the third type, reverse mergers involving two evolved stars that have already formed a core, are favored among progenitors with the lowest final core masses as can be seen in the bottom panels. We also see in each panel that these mergers originate from stars with the lowest initial masses. This is because they typically result from binary systems where the primary is a star of intermediate-mass, too low to undergo core collapse in case it had an isolated evolution.  If the primary had been more massive it would have been more likely to explode and create a compact object, possibly disrupting the system and anyway unable to produce another core-collapse supernova in case of a merger. The helium core or even the white dwarf formed from the primary needs to be  engulfed by the evolving secondary that has gained mass beforehand.  Because of their complex evolutionary history, with many instances of mass gain and loss for the two components, this is the only channel for which we find a non-negligible contribution that leads to lower  $M_{\rm He,preSN}$ than if they were in isolation.


\begin{figure}[t]\center
\includegraphics[width=0.51\textwidth]{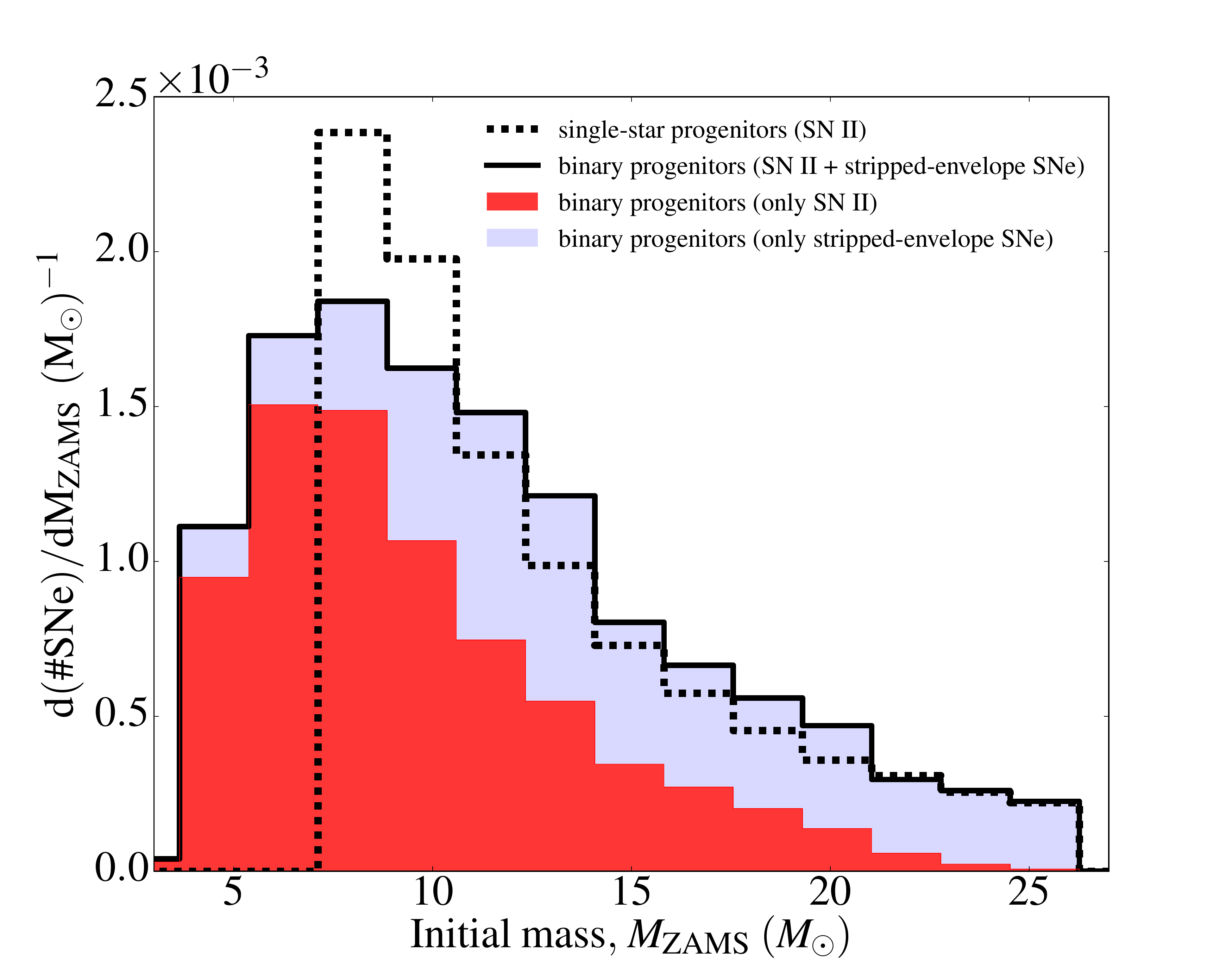}
\caption{Distribution of initial masses, $M_{\rm ZAMS}$, of SN progenitors with $M_{\rm ZAMS}< M_{\rm min,WR} \approx 27$ \Msun. For a population of single stars (dashed black line) we expect the distribution to follow the IMF, with all progenitors between $M_{\rm min,ccSN} \approx 7.6$  \Msun and $M_{\rm min,WR}$ leading to Type II SNe. The shape of the distribution changes for progenitors that experienced binary interaction (solid black line). The progenitors of SN~II arising from a population of binary interacting systems (red region) originate from  lower  on average $M_{\rm ZAMS}$, even below $M_{\rm min,ccSN}$.  In addition, many stars with $M_{\rm ZAMS}< M_{\rm min,WR}$ lead to stripped-envelope SNe due to binary interaction (blue region, stacked on top of the red region), avoiding a hydrogen-rich event. The results are for our standard assumptions and show the number of SNe per bin in  $M_{\rm ZAMS}$ per solar mass of star formation.
}
\label{ch6:fig:ZAMS_mass_dsitribution}
\end{figure}


\begin{figure}[t]\center
\includegraphics[width=0.51\textwidth]{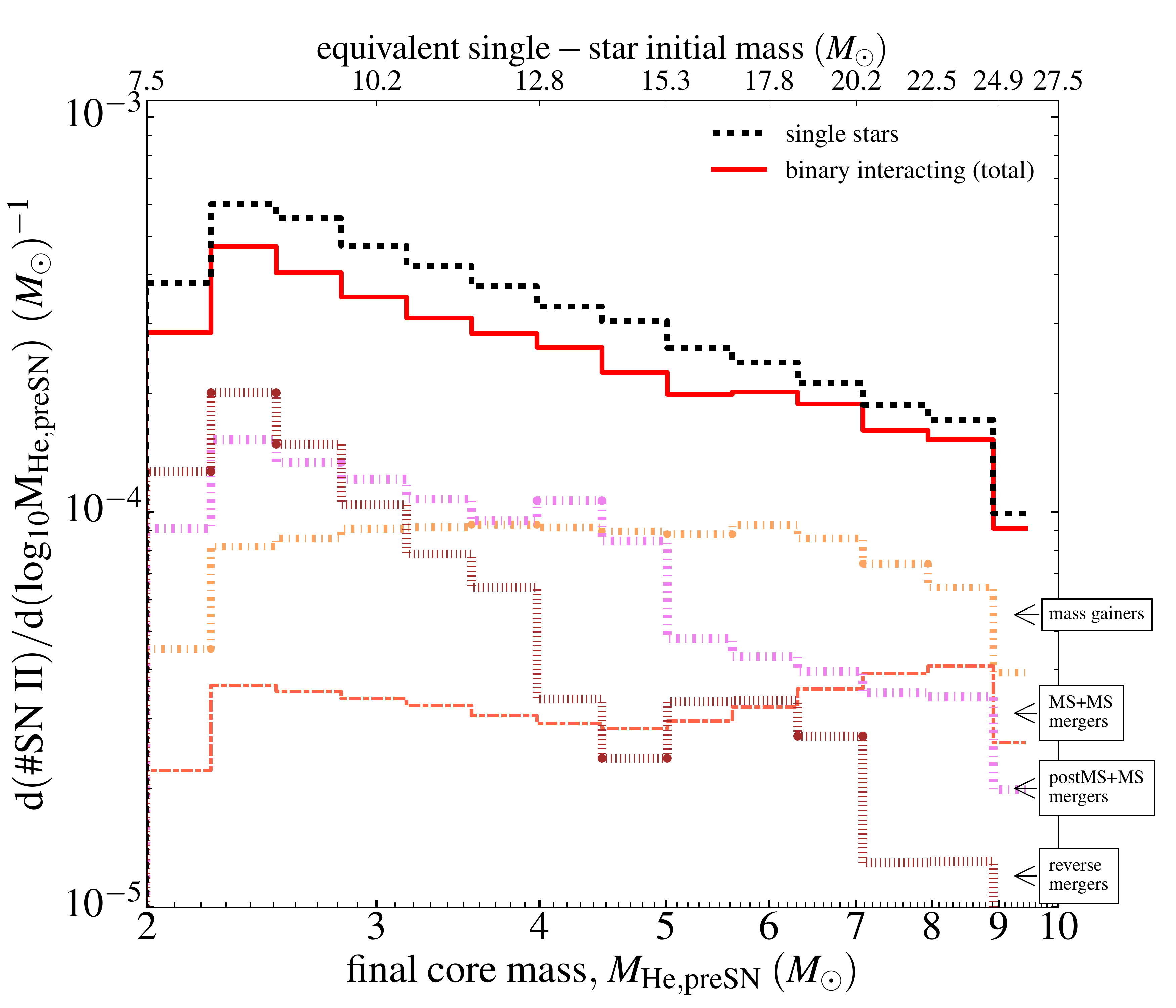}
\caption{Distribution of the pre-SN helium core mass of SN~II progenitors in case they experienced isolated evolution (black dashed  line) or have followed a channel of binary interaction (red solid line). 
The contribution of each different binary scenario is also shown (dashed and/or dotted lines). 
The equivalent single-star initial masses corresponding to each $M_{\rm He,preSN}$ (inverse of Equation~\ref{ch6:eq:helium_zams_singles}) is shown on the top x-axis.  
The results are for our standard assumptions and show the number of SN~II per logarithmic bin in $M_{\rm He,preSN}$ per solar mass of star formation.
} 
\label{ch6:fig:helium_slope}
\end{figure}

\subsection{Initial mass and final core mass distribution for Type II SN progenitors}\label{ch6:sec:results_slope}

Getting accurate parameters from the observations of individual SNe is challenging, but the sample of well-studied events is increasing. This means that we are starting to get constraints on the distribution of the characteristics of SN progenitors, including estimates of their initial and final core masses. We therefore analyze our simulations to obtain predictions for the distributions of initial masses and of final core masses for Type II SNe.

We show the initial mass distribution of SN~II in Figure~\ref{ch6:fig:ZAMS_mass_dsitribution}. For a population consisting only of single stars (black dashed line) the distribution is the initial mass function itself. All stars ranging between $M_{\rm min,ccSN} \approx 7.6$  \Msun and $M_{\rm min,WR}   \approx 27$ \Msun lead to a hydrogen-rich event.  The distribution of initial masses of progenitors of SN~II that experienced binary interaction (red, combining all the possible binary channels) is more spread and it is shifted toward lower initial masses.   A fraction of the progenitors originate from intermediate-mass stars, with  $M_{\rm ZAMS}<M_{\rm min,ccSN}$, which form a core massive enough to collapse only due to merging with their companion \citepalias{Zapartas+2017}. The distribution for binary progenitors of SN~II  for masses above  $M_{\rm min,ccSN}$ roughly follows the slope of the IMF  but is lower in number of events. This is partly because the distributions are normalized to the total mass of stars formed and the total mass of a binary system is by definition higher in each mass bin of $M_{\rm ZAMS}$ (which depicts the initial mass of the one of the two stellar components that finally explodes). 
 Another reason is that stars with $M_{\rm ZAMS}< M_{\rm min,WR}$ can lead to stripped-envelope SNe instead of of a Type II ones  if they are in binary systems (shown in blue, stacked on top of the red, in Figure~\ref{ch6:fig:ZAMS_mass_dsitribution}).  This is mainly due to mass stripping of the hydrogen-rich envelope of the primary star by its companion. Alternatively, for high initial masses close to $M_{\rm min,WR}$, the same can occur due to merging. If the merger product has eventually a strong enough wind mass loss to eject its hydrogen-rich envelope prior to explosion, it can avoid a Type II event. In summary, many stars in binaries in the mass range of $\sim 8-27$ \Msun produce a hydrogen-poor SN. At the same time the possibility of progenitors with initially masses lower than $M_{\rm min,ccSN}$ partially compensate to the number of Type II events. The overall effect is the shift of the distribution of initial mass of Type II SNe to lower values.

We also investigate the subsequent effect of binaries on the pre-SN core mass distribution, $M_{\rm He,preSN}$ . We show the results for our fiducial simulations for single and binary stars in Figure~\ref{ch6:fig:helium_slope} in black and red, respectively.  
On the top x-axis we also show the initial mass of a single star that would form a core as massive as shown in the bottom x-axis. To calculate this ``equivalent single-star initial mass'' we use the inverse of  Equation~\ref{ch6:eq:helium_zams_singles}, independent of whether the actual evolutionary path of the progenitor was influenced by binary interaction. 

The core mass distribution that we obtain for binary progenitors is somewhat shallower than for single stars. We fit the distributions with a power-law $dN/dM_{\rm He,preSN} \propto  M_{\rm He,preSN}^{\alpha_{\rm He}}$, 
which is represents a line of slope ($\alpha_{\rm He}+1$) in Fig.~\ref{ch6:fig:helium_slope}. 
We find a best fitting exponent $\alpha_{\rm He}$ of around $-2.0$  for single stars, while we find a lower value of around $-1.9$ for binary progenitors.  Both are shallower than the distribution of initial masses, which follows a standard IMF with a power-law slope of  $-2.3$ \citep{Salpeter1955,Kroupa2001}.   For single stars the difference is the effect of the relation between the initial and the final core mass (Equation~\ref{ch6:eq:helium_zams_singles}). For binaries the resulting distribution is the combined effect of the four dominant channels of binary mass gainers and mergers that contribute. The individual contributions are shown with dashed and dotted lines in Figure~\ref{ch6:fig:helium_slope} and are discussed below.

The distributions for mass gainers and for MS+MS mergers stand out because they are very shallow, nearly flat, and even rising at the high mass end in case of the MS+MS mergers.  These are also the channels in which the progenitor on average gains the largest amount of mass.   In our fiducial simulations, these are the two channels that contribute most significantly to the production of Type II SN progenitors with high final core masses, as we have already shown in section~\ref{ch6:sec:results_role_channels}, see the top panels in figure~\ref{ch6:fig:ZAMS_mass_forbinnedhecoremass_typeII}. These two channels are thus the most effective in enabling stars with lower initial masses to end their lives appearing as if they had been initially much more massive. They are the main reason that the final core mass distribution of SN~II is flatter for binaries, even though they originate on average from lower initial masses (red in Figure~\ref{ch6:fig:ZAMS_mass_dsitribution}). 

We see a different behavior for mergers of the second type (postMS+MS).   They have a slope that is roughly similar to the slope we obtain from our single stellar evolutionary models. The reason for this is that the merger occurs after the evolved star 
has formed a well-defined core which afterwards does not change significantly.   Thus, in our models this channel of merger has little effect on the final core mass distribution. 

Finally, reverse mergers show the steepest core distribution, peaking at  low final core masses. This is because they primarily originate from intermediate-mass binary systems where the primary star is close to or below $M_{\rm min,ccSN} \sim 7.6\Msun$ and does not explode prior to merging. Thus, they tend to form SN progenitors with final core mass equivalent to single stars with initial mass not much above the minimum mass for core collapse.

The slightly shallower helium core mass distribution when accounting for all binary channels implies a larger relative contribution of stars with massive final cores, as compared to a population of single stars with our assumed IMF.  One way to visualize this is by how flat the IMF would need to be for a purely single-star population to produce such a core-mass slope.  In Table~\ref{ch6:table:parameters_uncertainties} we compute the IMF slope, $\alpha_{\rm single\_IMF}$, that would have led a population of single stars to generate a final helium core-mass distribution that best fits the outcomes of each of our model binary populations. We further discuss these model variations in the next section, when comparing with observations.

\begin{figure*}[t]\center
\includegraphics[width=0.7\textwidth]{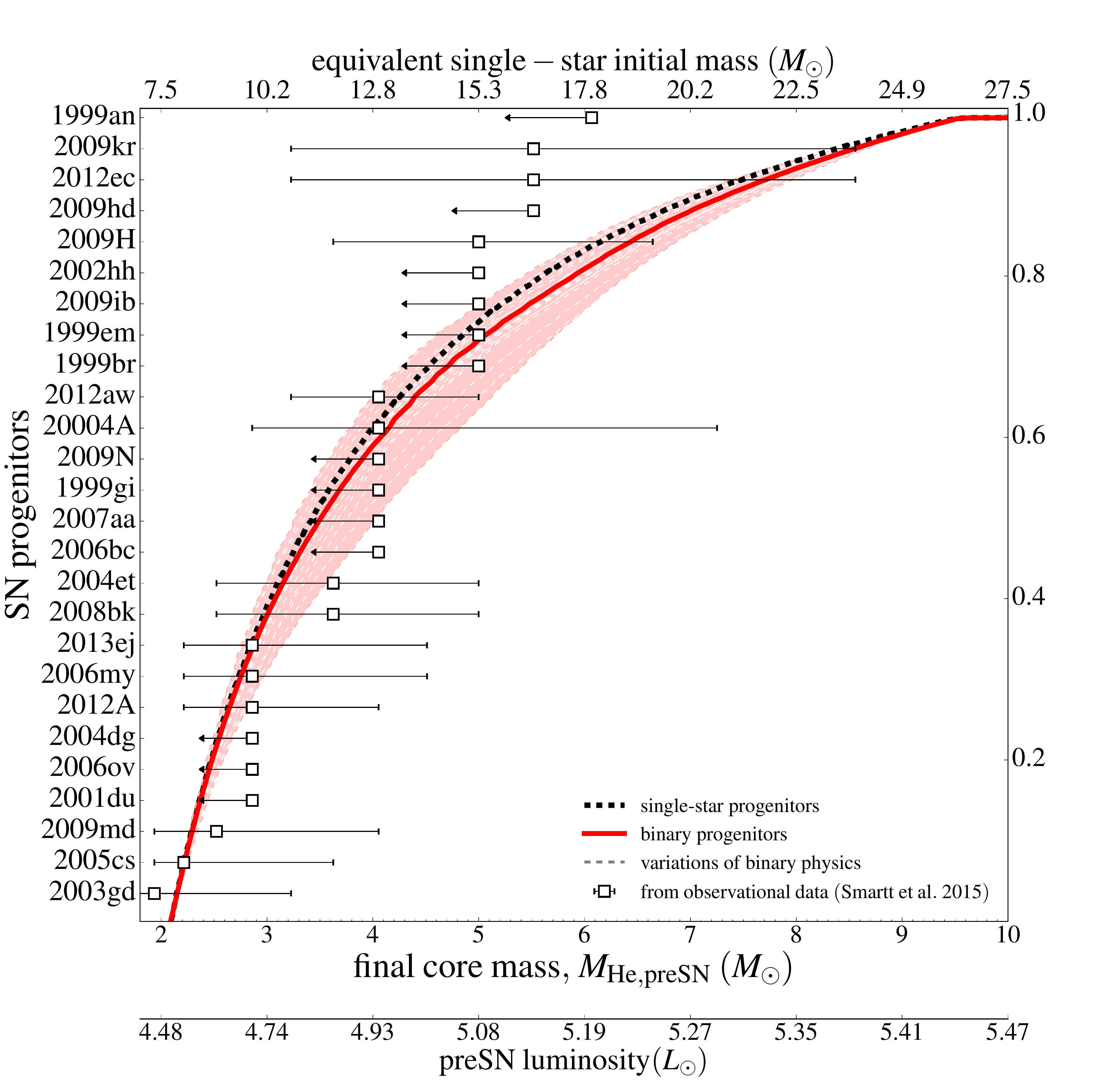}
\caption{Cumulative distribution of final core masses, $M_{\rm He,preSN} $, of Type II SN progenitors. White squares are the inferred core mass or upper limits of observed SN~II progenitors from \citet{Smartt2015}. 
For comparison we show our predicted cumulative distribution for progenitors that evolved as single stars (black dashed line) or including possible binary interactions (red solid line), following our standard assumptions. The equivalent single-star initial mass, $M_{\rm ZAMS}$, corresponding to each $M_{\rm He,preSN}$ 
is shown on the top x-axis.  
The equivalent pre-SN progenitor luminosity, corresponding to each $M_{\rm He,preSN}$ value according to single STARS models which were also used in \citet{Smartt+2009}, is shown on the bottom x-axis.  
We also show the range of uncertainty (red-shaded region with white dashed lines) of the expected distributions found for our simulated populations in which we vary the treatment of binary physics. These include among others the efficiency of mass transfer and of common envelope evolution, the initial period and mass ratio distribution and the binary fraction. 
}
\label{ch6:fig:cumulative_like_Smartt_binaries}
\end{figure*}

\section{Implications for interpreting direct detections of Type II SN progenitors and uncertainties} \label{ch6:sec:rsg_problem} 
\subsection{The sample of Type II SN progenitors \label{ch6:sec:sample}}

In this section we place our results in the context of observational constraints from direct progenitor detections. We discuss the implications for different observational techniques in section~\ref{ch6:sec:other_observational_constraints}.

Various groups have devoted large efforts to these searches, mostly using Hubble Space Telescope archival images \citep[e.g.,][]{Van-Dyk+1999,Smartt+2001, Van-Dyk+2003, Maund+2005a, Smartt+2009, Kochanek+2012,Davies+2018}. \citet{Smartt2015} made a compilation of SN progenitor observations. 
After briefly describing this sample here, we compare with our predictions in  section~\ref{ch6:sec:fiducial_comparison}, and discuss the effects of model variations and potential physical implications in sections~\ref{ch6:sec:comparison_binary_variations} and ~\ref{ch6:sec:comparison_other_variations}. 

We use the sample of nearby Type II SNe compiled by \citet{Smartt+2009} and \citet{Smartt2015}. This is a time- and volume-limited sample of SNe for which searches for progenitors were feasible. It contains 26 Type II SN events.  For each event high-resolution images taken before the explosion allowed direct detection of the SN progenitor, or could be used to calculate upper limits on the progenitor's luminosity.  \citet{Smartt+2009} compared the photometric measurements (or upper limits) with predictions for the pre-explosion luminosities from single stellar models by \citet{Eldridge+2004}.   

The final core masses for their best fitting progenitor models are not quoted in the paper by  \citet{Smartt+2009} directly, but they can be readily reconstructed using their Eq.~1.  The result of this is shown as square symbols in Figure~\ref{ch6:fig:cumulative_like_Smartt_binaries}, where the horizontal error bars indicate confidence intervals. This diagram is the analog to Figure~6 in  \citet{Smartt2015}, except that we chose to replace the initial mass with the final core mass on the bottom horizontal axis.  
The latter has the important advantage that it is more directly linked to what is inferred from pre-SN observations, circumventing uncertainties in the initial-final core mass relation. This is particularly important when considering binary progenitors, for which there is no unique mapping between the initial and final core mass, as we showed in section~\ref{ch6:sec:results_helium_zams_relation}.   

In the remainder of this section we discuss and overplot the helium core mass distributions derived from our simulations. When comparing our predictions with the observations, it should be kept in mind that the core masses of SN progenitors are inferred from the progenitor luminosity using single-star models.  We are mostly confident in this approach for the binary progenitors which experience interaction in an early evolutionary stage, which are expected to continue their further evolution in a fashion that is similar to that of a single star.  At present it is unknown exactly how accurate the estimates are for progenitors that result from more complex binary channels; our fiducial simulations suggest a contribution of 11\% for reverse mergers, and 14\% for mergers involving one postMS star, with both paths probably leading to more complex evolution than stars in isolation. A subset of these may lead to pre-explosion structures that differ from those that can be achieved by stars that evolve in isolation, for example  those that end their lives as blue supergiants \citep[e.g.,][]{Podsiadlowski1992,Justham+2014,Menon+2019}.  Blue supergiant SN progenitors do not follow even an approximate core-mass-luminosity relation, unlike red supergiants (see, e.g., Figure 3 of \citealt{Justham+2014}; for red supergiants see also, e.g., \citealt{Farrell+2020}). Overall we expect that this method should at least be reasonably reliable for stars which die as red supergiants, as in this observational sample.

This complexity of possible binary interaction of the progenitor is added on top of other sources of uncertainties.  
 In part there are caveats that are inherent to the observations and the possible role of selection effects, such as dust extinction of detected red supergiant progenitors \citep[e.g.,][]{Walmswell+2012,Beasor+2016} or limited accuracy of the exact position of the progenitor in the Hertzsprung-Russel diagram \citep{Davies+2018}. 
There are also uncertainties in the link between the inferred pre-SN mass of the progenitor star and its ZAMS mass that exist anyway in single stellar evolution, including the mass-loss rate through winds \citep[e.g.,][]{Smith2014, Renzo+2017}, the overshooting of the convective core \citep[e.g.,][]{Ribas+2000,Claret2007a,Brott+2011,Straniero+2019} and the theoretical preSN luminosities of the progenitors \citep[e.g.,][]{Farrell+2020}.

\subsection{Comparison with our fiducial models \label{ch6:sec:fiducial_comparison}}

In Figure~\ref{ch6:fig:cumulative_like_Smartt_binaries}  we overplot the cumulative final core mass distributions for hydrogen-rich SN progenitors from our simulations.   In black we show the results for our simulations for single stars and in red we show the distribution for our fiducial simulation for binary star progenitors.  
 Comparing the black and red line, we see that accounting for binary interaction does not have a large effect on the distribution. The distribution for binary progenitors is slightly shallower than that for single stars, as we showed in section~\ref{ch6:sec:results_slope}.

The data roughly follows the theoretical distribution for the lower mass end, perhaps depending on the interpretation of the upper limits, but one could argue that there is a deviation between the model predictions and the data at the high mass end.  The observed distribution appears to be steeper (or alternatively to have a lower maximum mass)  than predicted by our models.   In other words, our models over-predict the number of hydrogen-rich stars which reach core collapse with high helium core masses, when compared to the observed sample of Type II SN progenitors.  This is the same discrepancy as noted by \citet{Smartt+2009}, who coined the term ``red supergiant problem'' for the apparent lack of detections of Type II SNe resulting from progenitors with an initial mass higher than $\approx 16.5 \Msun$.    This corresponds to a helium core mass of  $\sim 5.75\,\Msun$  
(via Eq.~1 of \citealt{Smartt+2009}).  

Our simulations show that the inclusion of the effects of binary interaction does not help to solve or alleviate the claimed red supergiant problem. If anything, for our standard set of assumptions it makes it slightly worse.  To quantify this effect (still focusing only on stars which retain their hydrogen-rich envelope), we calculate the fraction of potential Type II SN progenitors, $R_{\rm massive}$, with a core more massive than $5.75\Msun$. Our models including binary systems predict a slightly larger $R_{\rm massive}$, about $0.2$, compared to $0.18$ for a single-star population (Table~\ref{ch6:table:parameters_uncertainties}).     If stars with core masses above that value result in ``failed'' events directly collapsing onto a black hole (as suggested in \citet{Smartt+2009} as a possibility and further discussed in section \ref{ch6:sec:comparison_other_variations}), $R_{\rm massive}$ would represent the fraction of hydrogen-rich stars at core collapse that would not produce a SN event. 

An alternative way to compare our models to the observational sample is to calculate the probability for each model population, $P_{\rm 0,massive}$, that in an observed sample of 26 SN~II progenitors, none would have been found with a helium core more massive than $5.75~\Msun$.  That is the same as the chance of picking randomly 26 SN progenitors from each model population (which is weighted according to initial conditions and taking into account stellar and binary evolution) and finding zero stars with equivalent single-star initial mass higher than $16.5 \Msun$. These are presented in Table~\ref{ch6:table:parameters_uncertainties} for all our simulations. Even given the caveats above, we consider this to be a helpful indication.  The probability of zero progenitor detections with helium core masses above $5.75~\Msun$, in a sample of 26 trials, is roughly a factor of two lower for our fiducial binary model than for purely single stars.

It is important to note however that this probability calculation is still based on the mass plane of the SN progenitors (of their cores in particular). \citet{Davies+2020} recently, by redoing the analysis of the claimed lack of high-mass progenitors from the point of view of observed red supergiant luminosities, argue that it may be an effect of the small sample size  and the steep luminosity distribution of the SN progenitors, and thus of lower statistical significance. In our work we refrain from directly predicting pre-SN luminosity distributions due to all the extra theoretical and observational uncertainties that our results would be subjected to, as discussed in section~\ref{ch6:sec:sample}.

\subsection{Impact of assumptions concerning the treatment of binary physics  \label{ch6:sec:comparison_binary_variations}}

Our predictions are subject to several uncertain assumptions. Here we discuss how variations in these assumptions affect the final core-mass distribution.  For a full list of the model variations that we considered, see Table \ref{ch6:table:parameters_uncertainties}.  For further discussion of the model variations we refer to \citetalias{Zapartas+2017} and \citetalias{Zapartas+2019}.

Our results are not very sensitive to assumptions concerning the treatment of binary physics or the initial distribution of mass ratios and orbital periods.   This can be seen from the red-dashed region in Figure~\ref{ch6:fig:cumulative_like_Smartt_binaries}, which shows the collective area spanned by model variations (these are listed in Table \ref{ch6:table:parameters_uncertainties} as the first group). Each variation in binary physics assumptions is shown with white dashed lines inside the region. Most of the simulations result in shallower distributions for the final core mass than our results for single stars.  None of the variations considered here solves the discrepancy between observations and the models at the high mass end.   For these variations, the probability of not detecting a more luminous progenitor than in the \citet{Smartt2015} sample remains below $P_{\rm 0,massive} \lesssim 10^{-2}$, typically close to $P_{\rm 0,massive}\approx 10^{-3}$, as for our fiducial model and purely single-star populations (also  shown in Table \ref{ch6:table:parameters_uncertainties}).

Simulations that predict a steeper cumulative core mass distribution are variations where we adopt  extreme assumptions.  These include a model variation where we assume that no mass is accreted during Roche-lobe overflow (Model 01), which prevents secondary stars from gaining mass, and a model variation in which we ignore the natal kick of compact objects (Model 10), which drastically reduces the number of binary systems that are disrupted upon the first SNe.  Such disruption is needed to prevent the secondary from being stripped by the primary star and therefore reduces the number of secondaries that end their lives as a Type II SN.   Finally, considering low efficiencies for common-envelope ejection (Models 12--14) leads to an increase in the contribution of reverse mergers, which dominate our predicted population of Type II SNe events with low final core masses. This also results in a steeper distribution in final core masses. 
The model variations that lead to the shallowest distributions include  those where we assumed a very high accretion efficiency (Model 03) or high efficiency for common envelope ejection preventing the contribution of mergers (Models 15 to 17).

\begin{figure*}[t]\center
\includegraphics[width=0.497\textwidth]{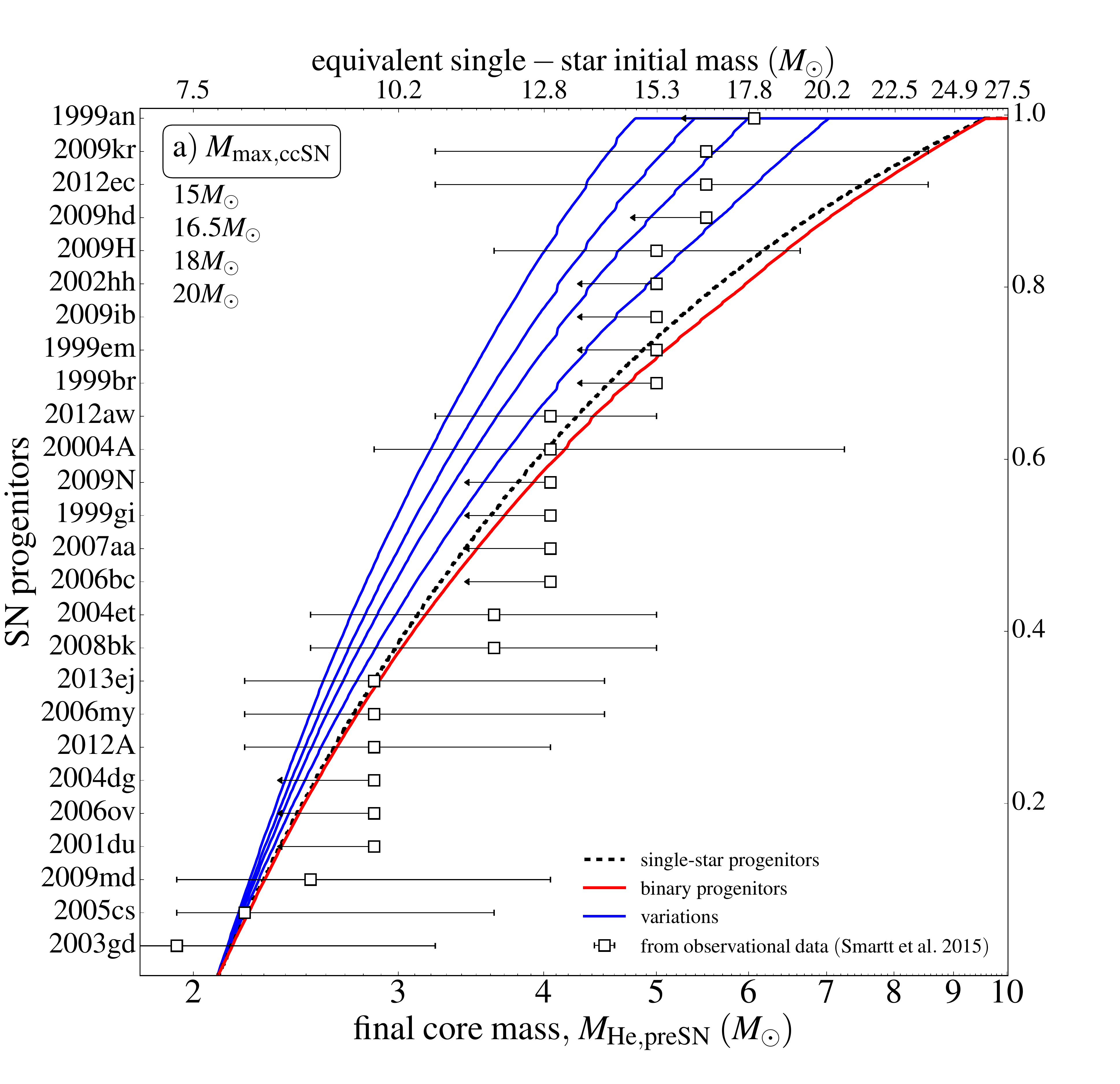}\includegraphics[width=0.497\textwidth]{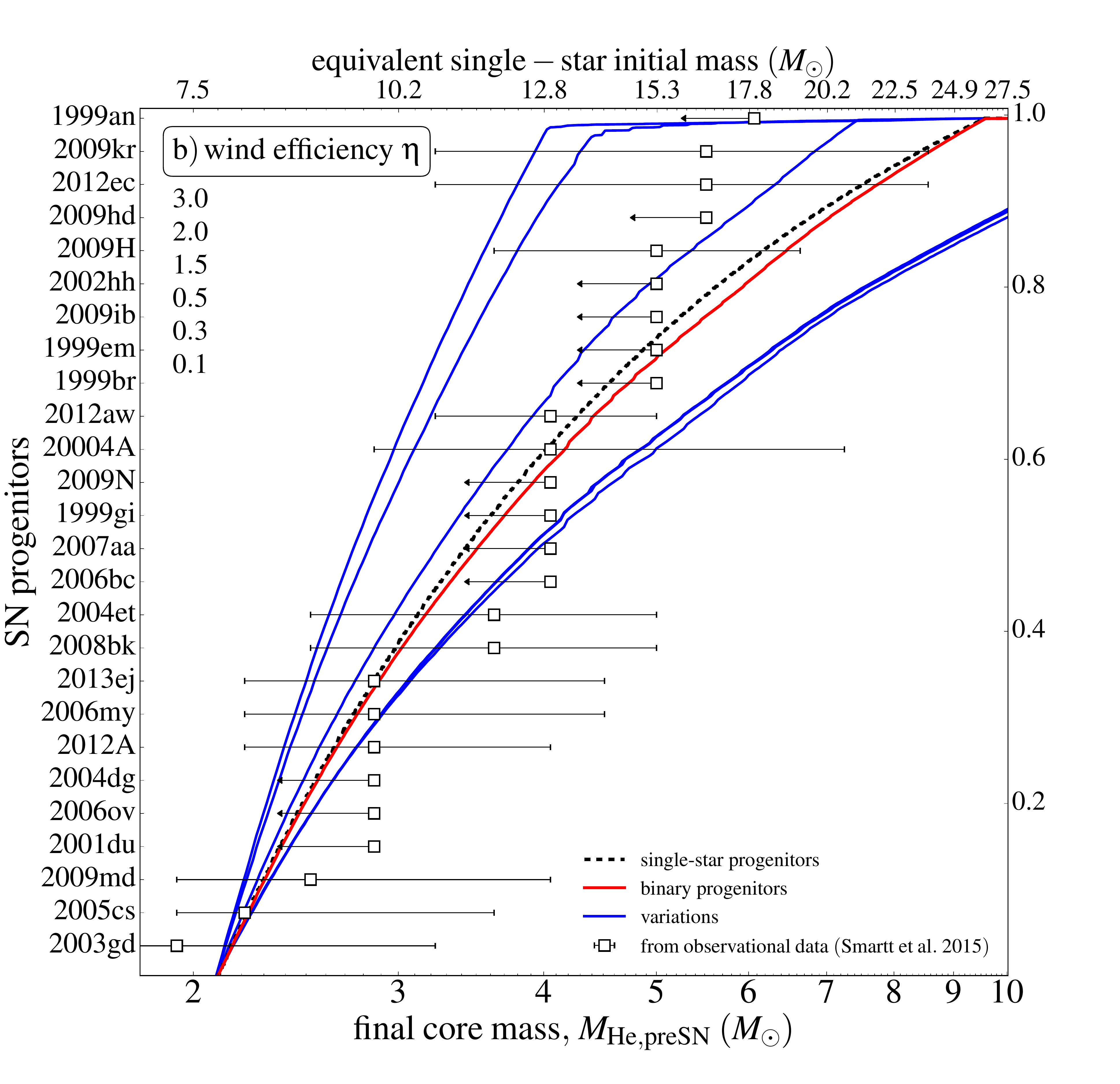}\\
\includegraphics[width=0.497\textwidth]{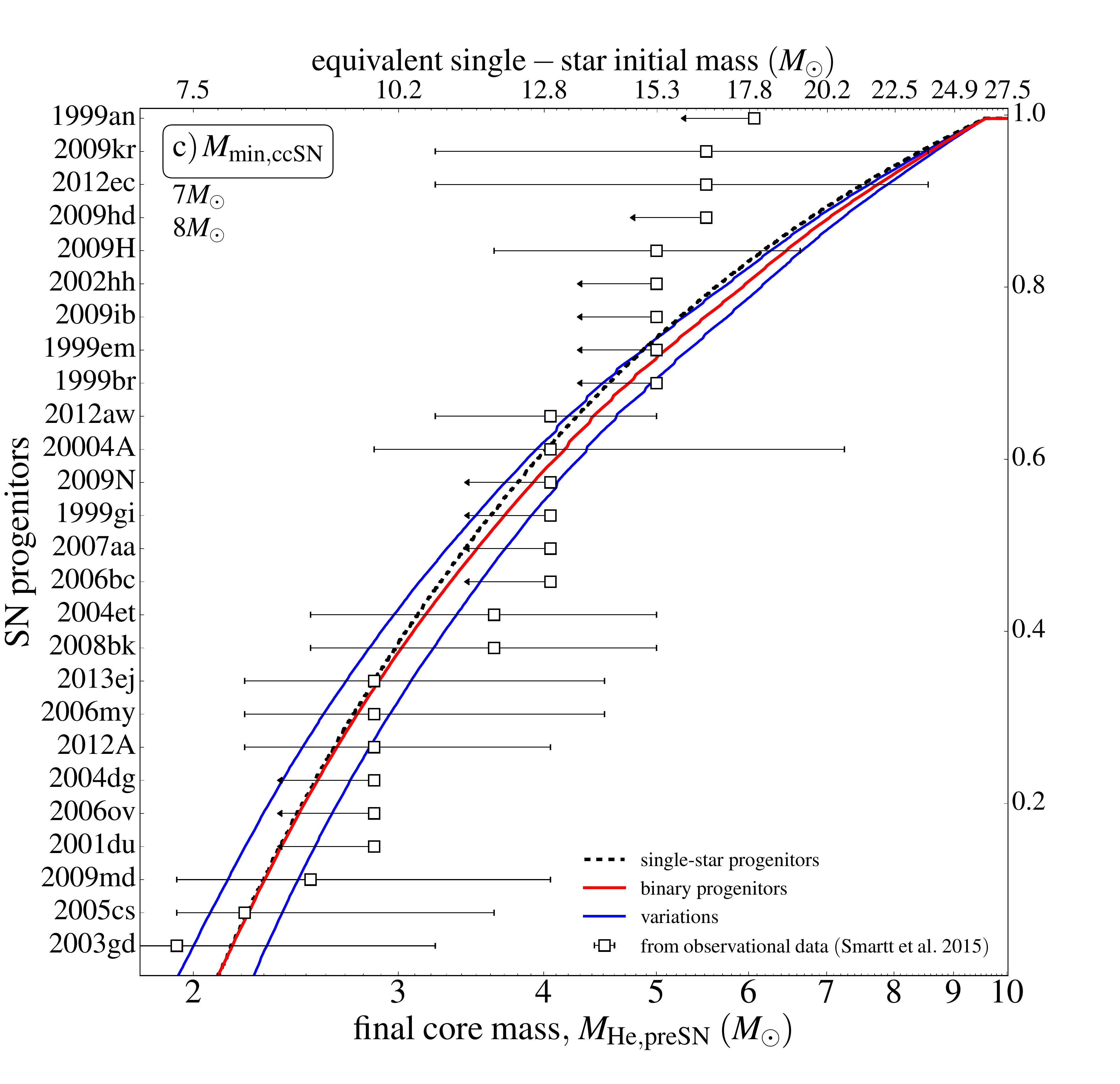}\includegraphics[width=0.497\textwidth]{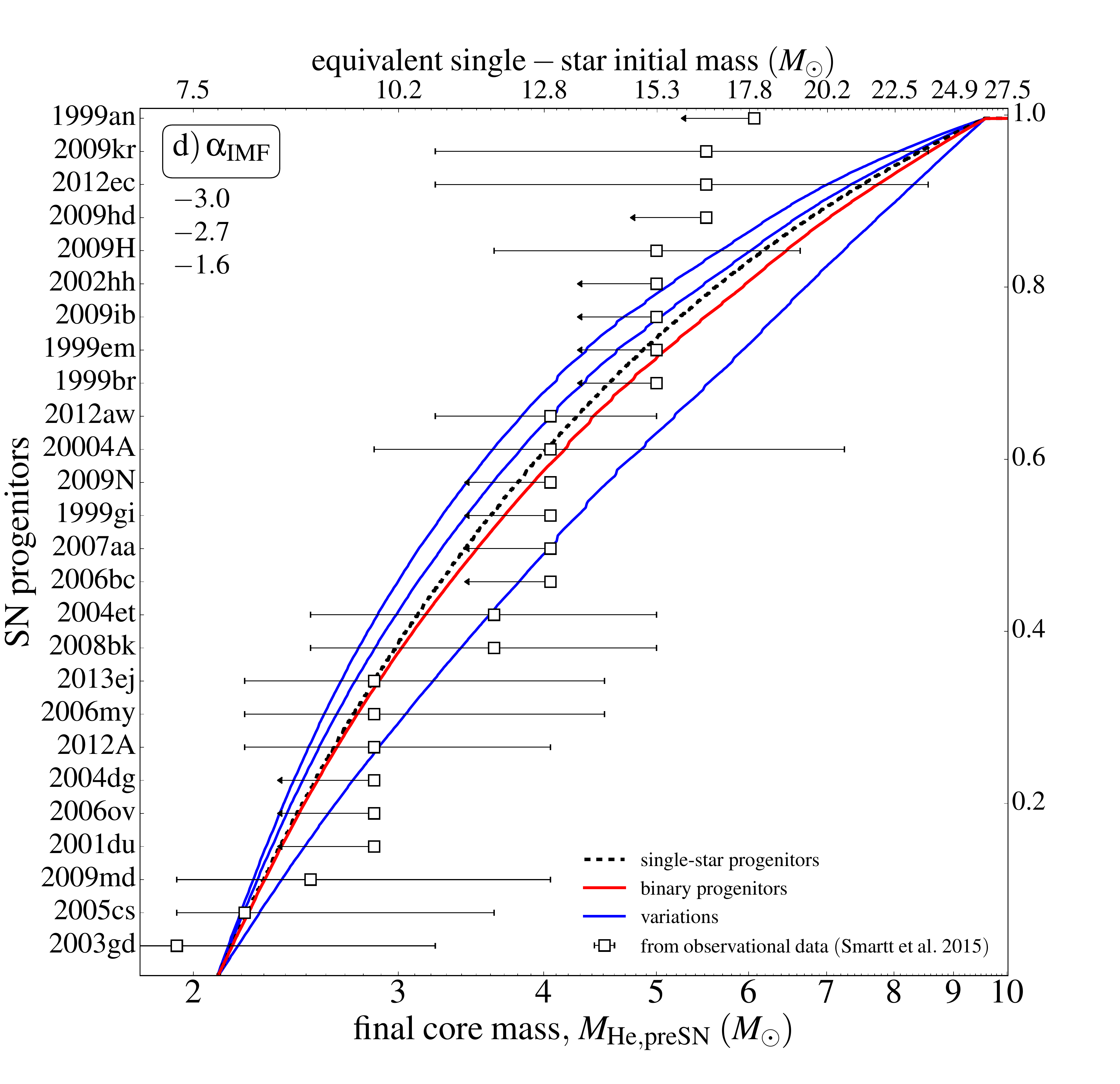}
\caption{Same as Figure \ref{ch6:fig:cumulative_like_Smartt_binaries} (with logarithmic x-axis here), but showing the results for other varying assumptions (blue lines): a) the mass threshold for failed SN collapsing onto black holes $M_{\rm{max,ccSN}}$ (Models 52, 53, 54 and 29),  b) the wind mass loss efficiency $\eta$ (Models 25, 26 and 51), c) the minimum mass for SN, $M_{\rm{max,ccSN}}$ (Models 30 and 31) and d) the slope of IMF $\alpha_{\rm IMF}$ (Models 32 to 34). These physical parameters influence the whole population, both single stars and in binary systems. The list of values depicted in the upper left corner shows a sequence of the varying distributions (blue lines) spanning from top left to bottom right of each panel.
}
\label{ch6:fig:cumulative_like_Smartt_singles}
\end{figure*}

\subsection{Impact of failed SNe, minimum mass for core-collapse SNe, stellar winds and the initial mass function \label{ch6:sec:comparison_other_variations}}

Since the inclusion of binary interaction does not significantly alter the final core mass distribution, we investigate which other model assumptions can potentially alleviate the apparent discrepancy between the predicted distribution for SN~II and the observational data, known as the red supergiant problem.   

\paragraph{Failed SNe ---}
In the last decade, there has been a debate about whether some stars do not produce a transient event because of full collapse onto a black hole \citep{OConnor+2011,Piro2013,Ertl+2016}. This is linked with the details of the explosion mechanism \citep[e.g.,][]{Janka2012}, subsequently determining the compact remnant that they leave behind \citep[e.g.,][]{Sukhbold+2016}.  Furthermore, \citet{Gerke+2015, Adams+2017} reported the disappearance of a red supergiant and proposed that this is an example of such an event. The possibility of failed SNe was suggested by \citet{Smartt+2009} as a potential solution to the apparent red supergiant problem.

We thus consider the simple possibility that cores above a mass threshold lead to a collapse with no visible transient.  
 In the upper-left panel of Figure \ref{ch6:fig:cumulative_like_Smartt_singles} we show the effect of variations in which we exclude all SNe that result from a progenitor with a helium core more massive than $ M_{\rm He,preSN} =$ 4.8,  5.5,  6 and 6.8 \Msun. These core masses correspond to a single-star initial mass of $M_{\rm max,ccSN} \approx $ 15, 16.5, 18 and 20 \Msun, respectively, where we used Eq.~\ref{ch6:eq:helium_zams_singles} in this work (Models 52--54 and 29). 
The small differences in the final core masses compared to \citet{Smartt+2009} are due to the slightly different initial - final core mass relation for single stars in that study from our Eq.~\ref{ch6:eq:helium_zams_singles}. 
 Decreasing $M_{\rm max,ccSN}$ shifts the cumulative distribution toward lower final core masses, changing the shape of the predicted curve to be closer to the values inferred from the observational sample. 
This  can lead to probabilities $P_{\rm 0,massive}$ of order unity for not finding SN progenitors with more massive cores  than the observed sample, as we exclude high mass core stars from our progenitor population (see Table \ref{ch6:table:parameters_uncertainties}). 

Earlier we referred to the calculated fraction of hydrogen-rich stars at core collapse which produce failed SNe, $R_{\rm massive}$,  assuming that the helium core mass governs this outcome (Table \ref{ch6:table:parameters_uncertainties}). 
For most models we use a threshold helium core mass of $\sim5.75\, \Msun$ which corresponds to that of an initial mass of 16.5 \Msun for a single star, following \citet{Smartt+2009}. 
This fraction of failed hydrogen-rich core-collapse events is around 0.2 for both our population of single stars and our fiducial simulation, as well as for most of our model variations. This ratio is consistent with 
up-to-date results of surveys that monitor for observational evidence of this phenomenon \citep{Kochanek+2008,Gerke+2015,Adams+2017,Adams+2017a}. The value is only moderately influenced even for extreme assumptions in the IMF slope, wind efficiency and  metallicity.   Models 29 and 52--54 consider different values of this parameter, $M_{\mathrm{max,ccSN}}$, which naturally cause the predicted ratio to change.

\paragraph{Stellar-wind mass loss ---} 
Altering the efficiency parameter of wind mass loss, $\eta$, results in significant variation in our predictions (see the upper-right panel of Figure~\ref{ch6:fig:cumulative_like_Smartt_singles}; Models 25, 26, 5, 52, 53 and 54). For lower wind efficiencies, the cumulative distribution is becoming even shallower, increasing the discrepancy with observations. This is  because stars are less  affected by winds and thus even progenitors with $M_{\rm ZAMS} \gtrsim 30-35 \Msun$ are able to produce Type II SNe. The reason that these distributions have very small differences is that stars above $M_{\rm ZAMS} \gtrsim 38 \Msun$ reach the Humphreys-Davidson limit \citep{Humphreys+1979} and  get stripped of their hydrogen-rich envelope anyway, even for low values of $\eta$. 

Increasing the mass-loss rate by a factor of 1.5 to 3 (Model 26, 53, 54) leads to much steeper distributions  
due to the high wind mass-loss rates that cause progenitors above single-star masses of 20 - 13 \Msun, respectively, to lose their hydrogen-rich envelope. They end their lives as a stripped-envelope SN instead of a Type II SN. This also severely affects the relative rates of those subclasses \citepalias{Zapartas+2019}.  
Although the distribution for $\eta=1.5$ seems visually quite consistent with observations, it should be primarily seen as a test for extreme assumptions of high mass loss, because, if anything, wind mass-loss efficiencies may be lower than in our standard assumptions \citep{Smith2014}. 
Changing the metallicity (Models 39-44, not shown in Figure \ref{ch6:fig:cumulative_like_Smartt_singles}) has a similar effect, since this also modifies the wind mass loss rates \citep[e.g.,][]{Vink+2001,Mokiem+2007}.   
Our model predictions for a super-solar metallicity of Z=0.03 appear to provide a good match to the data.  
However, we have no indications, for example from the host galaxies, that the majority of the sample consists of SNe originating from metal-rich progenitors.

\paragraph{IMF slope ---} 
Varying the slope of the IMF (Models 32--34; bottom-right panel of Figure~\ref{ch6:fig:cumulative_like_Smartt_singles}) does not strongly affect the shape of the final core-mass distribution.  Even an extremely steep IMF (e.g., $\alpha_{\rm IMF} = -3.0$ of Model 34) only increases the probability of missing a higher-mass SN~II progenitor by chance to $ P_{\rm 0,massive} \sim 2\times10^{-2}$.

\paragraph{Minimum mass threshold for core-collapse SN ---} 
In models 30 and 31 (bottom-left panel of Figure~\ref{ch6:fig:cumulative_like_Smartt_singles}), we change the minimum core mass for core-collapse SNe.  
A lower minimum threshold leads to a distribution which rises earlier in mass and thus sightly decreases the expected contribution of high mass progenitors above the observationally established mass limit.

\vspace{1em}

The parameters described in this subsection govern the initial conditions and the evolution of the whole population, both single stars and binary systems.  The sensitivity of our results to these assumptions is almost independent of whether we include binary systems or not in our simulations. The impact of these parameters on the final core mass distribution requires further careful inspection but this is not the aim of this paper in which we focus on the consequences of binary interactions.

\section{Predicted discrepancy of inferred initial masses of Type II SN progenitors between different observational methods}\label{ch6:sec:other_observational_constraints}


It is still challenging to establish a direct link between a SN event and its progenitor star. To this end, significant work has been invested in alternative methods to probe the characteristics of the progenitors of Type II SNe, in addition to searching for the progenitor in pre-SN images. 

A way to directly constrain the SN progenitor is to study the late-time, nebular-phase spectra of the SN when the ejecta become optically thin \citep{Jerkstrand+2012}. This provides an estimate of the oxygen core mass of the stellar progenitor.  A few studies have implemented this method, finding again a lack of high-mass progenitors \citep[e.g.,][]{Jerkstrand+2014, Tomasella+2013}. This is consistent with the red supergiant problem inferred from direct progenitor detections, discussed in section~\ref{ch6:sec:rsg_problem} (although see \citealt{Anderson+2018} for a reportedly high-mass progenitor case in a low-metallicity environment).    
Studies of the SN light curve evolution, that primarily constrain the ejecta masses of the event, are in general agreement with the above studies \citep[e.g.,][]{Valenti+2016, Martinez+2019,Eldridge+2019}.

All the techniques mentioned above (including the pre-SN imaging of the progenitors) directly probe the state of the star very close in time or at the moment of explosion. 
These studies put important constraints on the final progenitor state, but we should note that they do not directly probe the initial mass of the SN progenitor. In fact, we suggest avoiding translating the SN progenitor properties on the initial mass plane. Firstly, because it propagates the uncertainties of stellar evolution from formation up to explosion \citep[e.g.,][see also section~\ref{ch6:sec:sample}]{Davies+2020,Farrell+2020}. And especially because, in the case of binary interactions, the initial mass loses its predictive strength about the end result, due to the diversity of the different binary paths. We argue that the methods that estimate the pre-SN core mass and the stellar luminosity of the SN  progenitor 
are more closely related to what is defined as the final, pre-SN core mass in this study. Thus, if the possibility of binary mass gain or merging is not taken into account, we expect an overestimation of the inferred initial masses of the SN~II progenitor stars from these techniques (section~\ref{ch6:sec:results_helium_zams_relation}).

An independent way to probe the initial mass of the progenitor system is to infer the delay time between the formation of a star and its SN explosion, by studying the environment of the progenitor. This is achieved by effectively ``age-dating'' the parent population of the progenitor and thus inferring its initial mass. This method can mainly be applied to nearby SNe \citep{Murphy+2011,Williams+2014,Williams+2018}. As the local rate of SNe is low, the same technique has been used for SN remnants too \citep{Maoz+2010,Jennings+2014,Maund2017,Diaz-Rodriguez+2018,Auchettl+2019}.  
Several other studies investigate the association of SN~II with star-forming regions, through observations of nebular emission lines of the SN environment \citep[e.g.,][]{Anderson+2012,Habergham+2014, Xiao+2018, Kuncarayakti+2018,Schady+2019}, thus also providing an estimate of the initial mass of the progenitor star.  This technique has been also used to other stellar objects, like Luminous Blue Variables that may be progenitors of type IIn SNe. \citet{Smith+2015} and \citet{Aghakhanloo+2017} argue toward a binary interaction history of these objects to explain their relative isolation.

The lifetime of a star is not strictly determined by its initial mass as it may be affected by mass exchange with a companion \citepalias[\citealt{Crawford1955}, \citealt{De-Donder+2003},][\citealt{Xiao+2019}]{Zapartas+2017}. However, the initial masses of the stars in a binary determine the approximate timescale until binary interaction, which is usually the dominant component of the full time between formation and explosion. 
Thus, within uncertainties, the "age-dating" techniques estimate the actual initial masses of the SN progenitors (even the ones that experienced binary interaction). As we show in Fig.~\ref{ch6:fig:ZAMS_mass_dsitribution}, the expected initial mass distribution of SN~II progenitors is shifted to  lower initial masses \citepalias[and longer delay-times,][]{Zapartas+2017} than a canonical Salpeter IMF. Indeed, \citet{Jennings+2014} find a discrepancy in the inferred initial mass distribution of SN~II progenitors compared to a Salpeter IMF. They report that there may be a bias against the highest mass progenitors in their sample, 
but we argue that this result may be partially an outcome of the possible binary history of many of the progenitors.

\vspace{1cm}

\begin{figure}[t]\center
\includegraphics[width=0.52\textwidth]{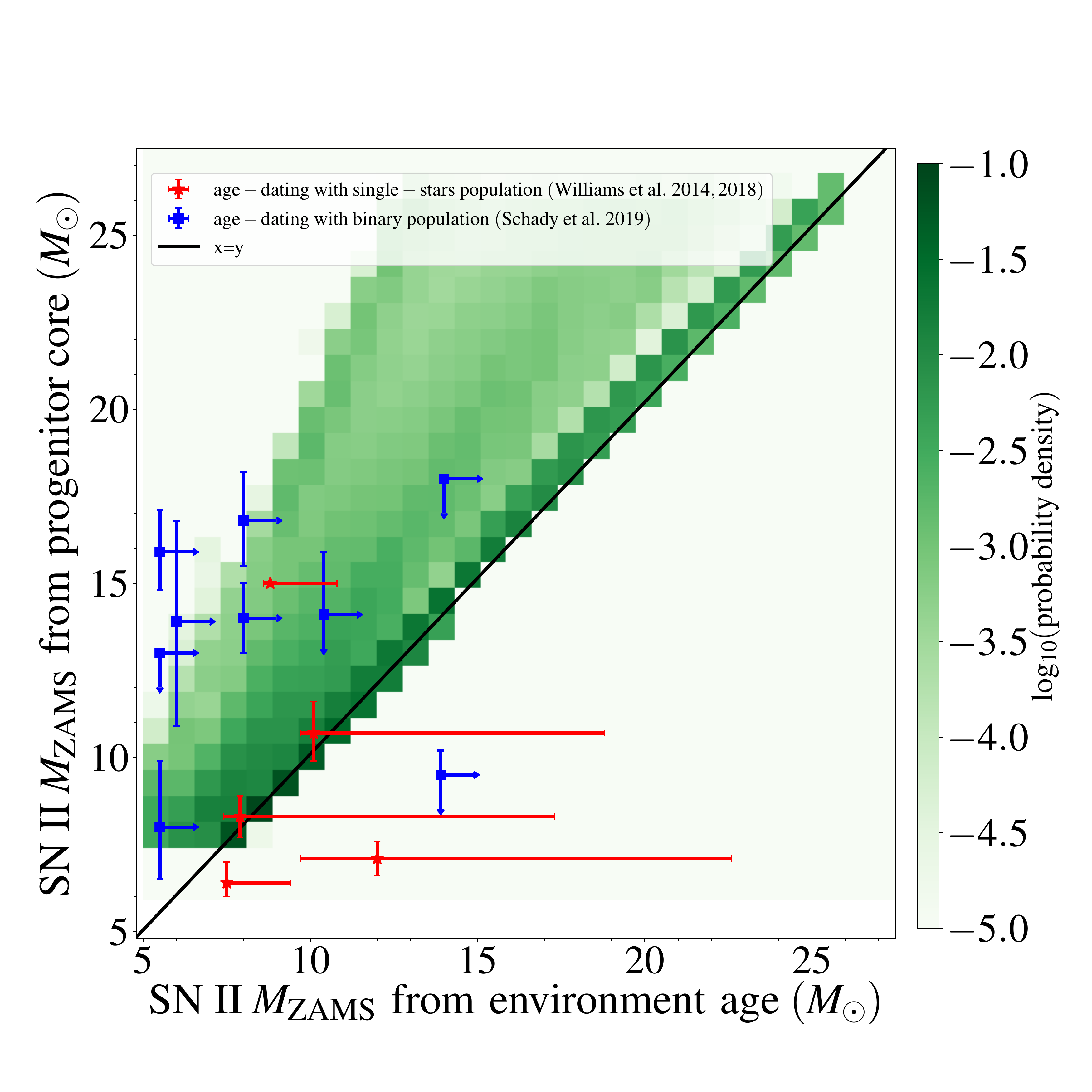}
\caption{Discrepancy in the inferred initial masses between methods that probe the pre-SN core mass or luminosity of the progenitor vs the ones that age-date its surrounding environment. Our results (2D probability density distribution in green) show the transformation of the simulated delay-time of the SN progenitor to $M_{\rm ZAMS}$ (x-axis), assuming a lifetime-mass relation for single stars \citepalias[Fig.~1 of ][]{Zapartas+2017}. The y-axis is the transformation of the simulated $M_{\rm He,preSN}$ to $M_{\rm ZAMS}$ using the inverse of Eq.~\ref{ch6:eq:helium_zams_singles}. The points are nearby Type II SNe with observed estimates of $M_{\rm ZAMS}$ with both groups of techniques, shown in Table~\ref{ch6:table:sn_sample}.
}
\label{ch6:fig:2D_discrepancy}
\end{figure}

Putting all these together, we argue that the two groups of observational methods discussed (the ones that study directly the SN progenitor and the others that examine its surrounding environment) probe different characteristics of the system that exploded and therefore  potentially infer different initial masses. In Fig.~\ref{ch6:fig:2D_discrepancy} we depict the theoretically expected 2D histogram of the inferred initial masses from the two techniques, for a realistic population of 50\% single stars and 50\% binary systems \citep[e.g.,][]{Sana+2012,Duchene+2013,Moe+2017}. We calculate the $M_{\rm ZAMS}$ that  age-dating methods would derive (x-axis) by computing the delay-time of the SN progenitors in our simulations and then transforming it to an initial mass of a single star that would have the same lifetime \citepalias[for details see Fig.~1 of ][]{Zapartas+2017}. 
At the same time, to compute the $M_{\rm ZAMS}$ that direct observations of the SN progenitors would infer, we transform the simulated final helium core to an initial mass, using the inverse of Eq.~\ref{ch6:eq:helium_zams_singles} in this study.

All single stars have the same value of $M_{\rm ZAMS}$ from the two methods (and are found on top of the $\mathrm{x}=\mathrm{y}$ line in Fig.~\ref{ch6:fig:2D_discrepancy}). However, we predict that for a significant fraction of SN~II, the initial mass of a progenitor inferred from age-dating the surrounding population is systematically lower than inferred from direct observational methods that study pre-SN properties. This is because a star may take long to explode due to its initial low mass but it has a more massive core at the moment of its explosion than corresponding to a single-star scenario of the same initial mass (due to merging or mass accretion during its lifetime). This discrepancy is potentially observable both for specific events and in statistical samples. 
%

 \begin{table*}[t]
  \caption{Sample of Type II SN progenitors with inferred initial masses from both directly probing the SN progenitor and age-dating the surrounding environment. \label{ch6:table:sn_sample}}
  \centering
   {\small
  \begin{tabular}{l||c|c|c}
   			& \multicolumn{2}{c}{environmental age-dating} & \\
           &  $M_{\rm ZAMS}$ assuming a &  $M_{\rm ZAMS}$ assuming a & $M_{\rm ZAMS}$ from\\
          SN event      &  single-star surrounding    & surrounding population of  & pre-SN core  \\
                     &population (\Msun)  & binary systems (\Msun)  & (\Msun)\\
\hline 
SN 1999br & (23.8)$^{(a)}$ & 10.4  $^{(b)}$     &  <$14.1^{+1.8}$ $^{(c)}$\\ 
SN 2001X &  (19.8)$^{(a)}$ & 8.0  $^{(b)}$  &       13-15$^{(d)}$\\
SN 2003dg & $7.5^{+1.9}_{-0.1}$ $^{(e)}$ &  ---    &  $6.4^{+0.6}_{-0.4}$ $^{(c)}$\\
SN 2004dg & (26.8)$^{(a)}$ & 14.0  $^{(b)}$ &     <$9.5^{+0.7}$ $^{(c)}$\\  
SN 2004et & $10.1^{+8.7}_{-0.4}$ $^{(e)}$ &  ---    &  $10.7^{+0.9}_{-0.8}$ $^{(c)}$\\
SN 2005cs & $12.0^{+10.6}_{-2.3}$ $^{(f)}$ &  ---    &  $7.1^{+0.5}_{-0.5}$ $^{(c)}$\\
SN 2006my & (18.5)$^{(a)}$ & 6.0  $^{(b)}$ & $13.9^{+2.9}_{-3.0}$ $^{(c)}$  \\
SN 2008bk & $12.0^{+10.6}_{-2.3}$ $^{(f)}$ &  ---    &  $8.3^{+0.6}_{-0.6}$ $^{(c)}$\\
SN 2008cn & (18.3)$^{(a)}$& 5.5  $^{(b)}$ & $15.9^{+1.2}_{-1.1}$ $^{(c)}$  \\
SN 2009H & (24.4)$^{(a)}$ & 14.0 $^{(b)}$ &   <18$^{(g)}$ \\
SN 2009N & (22.8)$^{(a)}$ & 5.5  $^{(b)}$&       <13$^{(g)}$ \\
SN 2009md & (18.2)$^{(a)}$ & 5.5 $^{(b)}$& $8.0^{+1.9}_{-1.5}$ $^{(c)}$ \\
SN 2012ec & (20.6)$^{(a)}$ & 8.0 $^{(b)}$ & $16.8^{+1.4}_{-1.3}$ $^{(c)}$  \\ 
SN 2017eaw & $8.8^{+2.0}_{-0.2}$ $^{(f)}$ &  ---    &  $\sim$ 15$^{(h)}$\\
  \end{tabular}
  }
  \tablefoot{\\
\tablefoottext{a}{$H \alpha$ emission by \citet{Schady+2019}, similar value to \citet{Kuncarayakti+2018}. Should be considered not valid\\ according to \citet{Schady+2019}, thus are not shown in Fig.~\ref{ch6:fig:2D_discrepancy}.}\\
\tablefoottext{b}{$H \alpha$ emission by \citet{Schady+2019}, should be considered a lower limit.}\\
\tablefoottext{c}{pre-SN image by \citet{Davies+2018}.}\\
\tablefoottext{d}{nebular line modeling by \citet{Silverman+2017}.}\\
\tablefoottext{e}{Surrounding resolved stellar population by \citet{Williams+2018}, possibly an upper limit.}\\
\tablefoottext{f}{Surrounding resolved stellar population by \citet{Williams+2014}, possibly an upper limit.}\\
\tablefoottext{g}{pre-SN image by \citet{Smartt2015}.}\\
\tablefoottext{h}{pre-SN image by \citet{Van-Dyk+2019}.}\\
}  
\end{table*}

We compare our results with a sample of Type II SNe that have estimates of their initial masses with both methods. The sample can be found in Table~\ref{ch6:table:sn_sample}. These events have estimates of their $M_{\rm ZAMS}$ based on pre-SN imaging of their progenitor \citep{Smartt2015,Davies+2018,Van-Dyk+2019} or modeling of the SN nebular spectrum \citep{Silverman+2017}. We made the argument that these techniques probe more closely the pre-SN core (corresponding to the y-axis in Fig.~\ref{ch6:fig:2D_discrepancy}). At the same time, these SNe have independent estimates of their $M_{\rm ZAMS}$, based on studies of their surrounding environment that constrain the age of the progenitors (corresponding to the x-axis in Fig.~\ref{ch6:fig:2D_discrepancy}).

Part of the sample has been presented in \citet{Schady+2019}, where they used $H\alpha$ emission lines in the surrounding region as an age tracer. They show that this method is highly sensitive to the modeling of the stellar environment, because binary products make a region look younger than its actual age.  This can lead to a possible underestimation of the delay-time and thus an overestimation of the initial mass of a SN~II progenitor. 
\citet{Schady+2019} address the reported high initial masses derived in \citet{Kuncarayakti+2018} where single-stellar populations have been assumed for the age-dating (2nd column in Table~\ref{ch6:table:sn_sample}, in parenthesis), arguing toward lower initial masses instead if binarity is taken into account (3rd column in Table~\ref{ch6:table:sn_sample}).  
In Fig.~\ref{ch6:fig:2D_discrepancy} we show their derived values 
where the full effect of binarity on the time evolution of $H\alpha$ emission lines is taken into account  (blue squares in Fig.~\ref{ch6:fig:2D_discrepancy}). They argue that these values of $M_{\rm ZAMS}$ should be considered lower limits because possible photon leakage decreases the effect of binarity in the age-dating via $H\alpha$ emission lines \citep{Xiao+2018}.   
The observed values of $M_{\rm ZAMS}$, where the effect of binarity in the age-dating is considered, are roughly in-line with our prediction of the discrepancy between the two groups of observational methods. In fact, for some events the initial masses from studying the environment are much lower than our theoretical predictions, although they should be considered lower limits anyway.

\citet{Williams+2018} have also compared initial mass estimates from the two different method groups for a sample of SN events. We include these events in our sample too, with updated  $M_{\rm ZAMS}$ estimates from pre-SN images by \citet{Davies+2018} and \citet{Van-Dyk+2019} for most of them. \citet[][adding to previous work from \citealt{Williams+2014}]{Williams+2018} derive an estimate of the age of the region by modeling the resolved stellar population around the SN (2nd column in Table~\ref{ch6:table:sn_sample}, not in parenthesis). They report a broad agreement between their initial mass estimates and the ones from pre-SN images (red stars in Fig.~\ref{ch6:fig:2D_discrepancy}). So, there is no obvious trend for SNe that show a discrepancy between these two methods, although SN2017eaw is a clear exception. \citet{Williams+2014,Williams+2018} used single stellar modeling, not taking into account the effect of binarity on the observed color-magnitude diagram of a stellar population. Binaries would lead to hotter and more luminous stars, potentially leading to underestimations of a population's age \citep[e.g.,][]{Eldridge+2009, Schneider+2015}. At the same time, they considered only the recent star-formation history of a region up to $50$ Myrs before the SN, based on the maximum lifetime of a single star progenitor. \citet{De-Donder+2003} and \citetalias{Zapartas+2017} show that we should expect even higher delay-time for SN progenitors originating from intermediate-mass stars. Thus, allowing for the possibility of longer delay-times than 50 Myrs may result in  lower derived initial progenitor masses. It would be interesting to see the results if those effects are considered, as both of them qualitatively are expected to lower the $M_{\rm ZAMS}$ estimates of these studies.

Although the theoretical and observational uncertainties in both groups of methods are significant, it is interesting that some observational points are found above the x=y axis. As a case study, we can discuss the  significant discrepancy of the SN 2017eaw initial mass values. Here we used $M_{\rm ZAMS} \sim 15$ \Msun derived by \citet{Van-Dyk+2019} from pre-SN images, but \citet{Kilpatrick+2018} and \citet{Rui+2019} also  infer an initial mass around $12-13$ \Msun. Interestingly, all of them are higher than the estimate of $8.8^{+2.0}_{-0.2}$ from age-dating of \citet{Williams+2014}. Conclusive statements  are difficult to make without strong constraints from both of the groups of techniques, but a high discrepancy in the values derived from these methods can potentially be an indicator of a binary history of a Type II SN progenitor. This indication will be even stronger in case it is combined with an exceptionally long  delay-time of the progenitor ($> 50 $ Myrs) which would correspond to a single star initial mass below the mass threshold for a collapse $M_{\mathrm{min,ccSN}} \sim 7.5 \Msun$. These late SNe can be a signature of binary interaction of the progenitor too  \citepalias{Zapartas+2017}.  A bigger sample of Type II SNe with accurate estimates of the pre-SN core or luminosity and, at the same time, with careful study of the age of their environment will shed more light on possible binary history of their progenitors.

On the other hand, it is interesting to discuss a few events that are found well below the x=y line, which is not expected in our findings. For example, the $9.5^{+0.7}$ \Msun  upper limit of  SN 2004dg in pre-SN images seems to be lower than the one derived by age-dating, even when a surrounding population of binaries is considered. Given our theoretical expectations, a physical reason for this to occur would be that the progenitor was heavily obscured and therefore its inferred mass from the nondetection of it is an underestimate. Further studies of these group of unexpected events would be needed to shed more light on their evolutionary history. 

In our analysis we did not include the SNe reported in \citet{Williams+2014}, \citet{Williams+2018}, and \citet{Schady+2019} where the progenitor was fully or partially stripped of its envelope (Type Ib iPTF13bvn, and Type IIb's SN1993J and SN2011dh) because our predictions apply for hydrogen-rich, Type II SNe only. We also excluded SN1987A because single star models struggle to simulate its blue supergiant progenitor, thus they cannot confidently infer what its initial mass would be if it had somehow evolved from a single star.

We note that a potential complication for the methods of age-dating is the possibility that the SN progenitor was ejected from its initial binary system prior to its own explosion. This is exactly the case for SN~II progenitors that gained mass, according to our study. These stars can travel tens to hundreds of pc away from their birthplace until explosion  \citep{Eldridge+2011,Renzo+2019}. We expect a statistically weaker correlation of SNe from mass gainer progenitors to young, star-forming regions compared to the bulk population of SNe~II.  The relative spatial isolation of the progenitor of SN2017eaw \citep{Van-Dyk+2019} also fits to this possibility.

\section{Summary and conclusions}\label{ch6:sec:summary}

This work investigates the effect of binary interactions on the initial and final core mass distribution of hydrogen-rich Type II supernova progenitors. We expand on \citet{Zapartas+2019}, where we find that around 1/3 to 1/2 of Type II SN progenitors have experienced mass accretion or merged with a companion prior to explosion.

\begin{itemize}
\item We demonstrate that binary interactions 
lead to a broad relation between initial mass and final core mass because of the variety of possible evolutionary paths. Consequently, this breaks the tight relation between pre-explosion properties and initial stellar mass. 

\item We find that progenitors that have experienced mass accretion or merging with a companion, for a given helium core mass, typically originate from lower initial masses than would be expected from single-star evolution (Figs.~\ref{ch6:fig:2D_hecore_ZAMS} and \ref{ch6:fig:ZAMS_mass_forbinnedhecoremass_typeII}). We argue that pre-SN images of Type II SN progenitors give a more direct insight into the final helium core masses than  total masses, since the final stellar luminosity is more closely related to the former \citep[e.g.,][]{Giannone1967}.  
Thus, if one does not consider the possibility of a binary history for a Type II SN progenitor, it may often lead to overestimation of its inferred initial mass.

\item We find that the initial mass distribution of SN~II progenitors is broadly shifted to lower initial masses than expected from a canonical Salpeter IMF of single stars (Fig.~\ref{ch6:fig:ZAMS_mass_dsitribution}). Our fiducial binary population simulation produces a marginally shallower distribution of final helium core masses than from a population of purely single stars. This is best fit by a power-law slope of approximately -1.9, compared to about -2.0 for a single-star population (Fig.~\ref{ch6:fig:helium_slope}). 

\item Binary interactions do not appear to significantly affect the red supergiant problem, that is, the reported lack of SN progenitors with final inferred helium cores that correspond to initial single-star masses above $\sim16.5 M_{\odot}$ \citep[][]{Smartt+2009,Smartt2015}. This seems to hold for all our realistic variations in binary physics assumptions (Fig.~\ref{ch6:fig:cumulative_like_Smartt_binaries}).  

\item We also show the effect of  other physical processes and conditions that affect the final core distribution (Fig.~\ref{ch6:fig:cumulative_like_Smartt_singles}). The  inclusion of failed SNe or the increase of the typically used wind mass-loss rates   
result in a steeper distribution skewed toward lower core masses.

\item We argue that the initial masses inferred from the surrounding environments of Type II SN or SN remnants are typically closely associated with the actual initial primary-star mass of the progenitor, even for binary progenitors, as long as binarity is properly taken into account when age-dating the region.  On the other hand, the luminosity of a red supergiant progenitor would correlate most strongly with its final core mass, and inferences based on the pre-SN state would tend to overestimate the initial stellar mass of the progenitor if they ignore the possibility of mass gain or merging (section~\ref{ch6:sec:other_observational_constraints}). Thus, we predict the inferred initial masses from studies of the surrounding environment to be systematically lower than inferred in the techniques that estimate the pre-SN core or luminosity (Fig.~\ref{ch6:fig:2D_discrepancy}), consistent with the new observational analysis of \citet{Schady+2019}. An accurate and significant discrepancy in the values derived from these methods for a particular Type II SN may even suggest a binary history for the progenitor of that SN. 

\end{itemize}

We suggest that future studies of Type II supernova progenitor populations and explosion diversity should not assume a one-to-one mapping between final progenitor properties and initial stellar mass. This suggestion applies even where individual cases do not seem to require an explanation involving binary interactions.

\setlength{\tabcolsep}{2pt}
\begin{table*}[p] 
  \caption{Model variations of the assumptions that affect the binary population specifically as well as our whole stellar population. We show the slope of the distribution of final helium cores, $\alpha_{\rm He}$, as well as the corresponding IMF slope of a single-star population, $\alpha_{\rm single\_IMF}$. $R_{\rm massive} $ is the fraction of hydrogen-rich stars at collapse with cores more massive than $5.75\Msun$ that corresponds to a single star of 16.5 \Msun \citep{Smartt+2009}, which has been suggested lead to ``failed'' SNe. 
$P_{\rm 0,massive}$ is the possibility of having zero progenitor detections of these high core mass progenitors out of 26 observational trials by chance. In the first column we show the model number of each variation, which corresponds to a simulation of \citetalias{Zapartas+2017} in case it is not in parenthesis.
\label{ch6:table:parameters_uncertainties}}
  \centering
   {\small
   \begin{tabular}{ll|l|l|l|l}
  \hline \hline 
   \\[0.01ex]
Model                                   &Description            & slope of   & slope of equivalent  & ratio of hydrogen-rich&$10^{3} \times$ prob. of \\
in \citetalias{Zapartas+2017} &                          & $M_{\rm He,preSN}$         & single-star    & progenitors with  & 0 high mass \\
 &                          &   distribution,      &  IMF distribution,   & high mass cores, & detections,\\
  &                          &   $\alpha_{\rm He}$      &  $\alpha_{\rm single\_IMF}$    & $R_{\rm massive} $ & $P_{\rm 0,massive} \times 10^{3} $ \\
\hline
 \\[0.0015ex] 

{\bf --} &  {\bf FIDUCIAL POPULATION      }  (standard set of assumptions)                           &         {\bf  -1.89} &     {\bf -2.20}&    {\bf 0.2} & {\bf 4.6}\\
[1ex]
 -- &  population of only single stars (standard set of assumptions)                            & -2.01 &  -2.35 &   0.18 &     9.0 \\
[1.5ex]
\multicolumn{5}{l}{\it \bf Variations of assumptions affecting the binary population} 
\\ [0.01ex]

\multicolumn{5}{l}{\it  -- Parameters of stable mass transfer} 
\\[0.01ex]
  01 &  accretion efficiency $\beta=0$                          &   -2.22 &  -2.62 &   0.16 &     12.6 \\
  02 &  accretion efficiency $\beta=0.2$                        &   -2.04 &  -2.38 &   0.18 &     8.6 \\
  03 &  accretion efficiency $\beta=1$                          &   -1.34 &  -1.50 &   0.26 &     1.0 \\
  04 &  angular momentum loss $\gamma=0$                        &   -1.80 &  -2.08 &   0.22 &     3.0 \\
  05 &  angular momentum loss $\gamma$:circumbinary disk        &   -1.91 &  -2.22 &   0.19 &     6.1 \\
 [1ex]
\multicolumn{5}{l}{\it --  SN kick} 
 \\[0.02ex]
  10 &  no natal kick in SNe, $\sigma_{\mathrm kick} = 0$       &   -2.25 &  -2.66 &   0.17 &     11.3 \\
  11 &  extremely high kick in SNe, $\sigma_{\mathrm kick} = \infty$ &   -1.88 &  -2.19 &   0.21 &     4.2 \\

\multicolumn{5}{l}{\it  -- Common envelope evolution} 
 \\[0.01ex]
  12 &  $\alpha_{\mathrm{CE}} = 0.1$                            &   -2.26 &  -2.66 &   0.16 &     12.4 \\
  13 &  $\alpha_{\mathrm{CE}}= 0.2$                             &   -2.12 &  -2.49 &   0.18 &     8.6 \\
  14 &  $\alpha_{\mathrm{CE}} = 0.5$                            &   -1.87 &  -2.17 &   0.21 &     4.0 \\
  15 &  $\alpha_{\mathrm{CE}}= 2.0$                             &   -1.86 &  -2.16 &   0.21 &     3.6 \\
  16 &  $\alpha_{\mathrm{CE}} = 5.0$                            &   -1.60 &  -1.83 &   0.24 &     1.6 \\
  17 &  $\alpha_{\mathrm{CE}} = 10.0$                           &   -1.47 &  -1.67 &   0.26 &     1.1 \\
  18 &  $\lambda_{\mathrm{CE}}=0.5$                             &   -2.02 &  -2.36 &   0.20 &     4.9 \\
\multicolumn{5}{l}{\it  -- Parameter space for unstable mass transfer and mergers} 
 \\[0.01ex]
  19 &  $q_{\mathrm{crit,MS}} = 0.25$                           &   -1.90 &  -2.21 &   0.20 &     4.5 \\
  20 &  $q_{\mathrm{crit,MS}} = 0.8$                            &   -1.90 &  -2.21 &   0.20 &     4.5 \\
  21 &  $q_{\mathrm{crit,HG}} = 0.0$                            &   -1.90 &  -2.20 &   0.20 &     4.6 \\
  22 &  $q_{\mathrm{crit,HG}} = 0.25$                           &   -1.83 &  -2.12 &   0.21 &     3.5 \\
  23 &  $q_{\mathrm{crit,HG}} = 0.8$                            &   -1.94 &  -2.26 &   0.18 &     8.6 \\
  24 &  $q_{\mathrm{crit,HG}} = 1.0$                            &   -1.95 &  -2.28 &   0.17 &     10.2 \\
  06 &  merger $\mu_{\mathrm{loss}}=0$                          &   -1.86 &  -2.15 &   0.21 &     3.8 \\
  07 &  merger $\mu_{\mathrm{loss}}=0.25$                       &   -1.87 &  -2.17 &   0.20 &     4.4 \\
  08 &  merger $\mu_{\mathrm{mix}}=0$                           &   -1.90 &  -2.21 &   0.20 &     4.5 \\
  09 &  merger $\mu_{\mathrm{mix}}=1$                           &   -1.89 &  -2.20 &   0.20 &     4.4 \\
 27 &  WD mergers exclusion if not massive companion           &   -1.75 &  -2.02 &   0.22 &     3.1 \\
\multicolumn{5}{l}{\it  -- Initial binary distributions} 
 \\[0.01ex]
  35 &  initial q distr. slope $\kappa=-1$                      &   -1.90 &  -2.20 &   0.20 &     5.1 \\
  36 &  initial q distr. slope $\kappa=+1$                      &   -1.87 &  -2.17 &   0.21 &     3.8 \\
  37 &  initial period distr. slope $\pi=+1$                    &   -1.97 &  -2.30 &   0.19 &     6.8 \\
  38 &  initial period distr. slope $\pi=-1$                    &   -1.88 &  -2.18 &   0.20 &     5.2 \\
\multicolumn{5}{l}{\it  -- Binary fraction} 
 \\[0.01ex]
45 &  $f_{\mathrm{bin}} = 0.3$                                &   -1.98 &  -2.32 &   0.18 &     7.6 \\
(55) &  $f_{\mathrm{bin}} = 0.5$                                &   -1.96 &  -2.28 &   0.19 &     6.6 \\
00 &  $f_{\mathrm{bin}} = 0.7$                                &   -1.93 &  -2.22 &   0.19 &     5.8 \\  
47 &  mass dependent binary fraction, $f_{\mathrm bin}(M)$    &   -1.96 &  -2.29 &   0.19 &     6.3 \\
 \hline \hline
\multicolumn{5}{l}{\it  -- (continued in next page)}
  \end{tabular}
  }
  \end{table*}

\setlength{\tabcolsep}{2pt}
\begin{table*}[p] 
  \centering
   {\small
   \begin{tabular}{ll|l|l|l|l}
  \multicolumn{5}{l}{\it  -- (continued from previous page)}\\
  \hline \hline 
   \\[0.01ex]
Model                                   &Description            & slope of   & slope of equivalent  & ratio of hydrogen-rich&$10^{3} \times$ prob. of \\
in \citetalias{Zapartas+2017} &                          & $M_{\rm He,preSN}$         & single-star    & progenitors with  & 0 high mass \\
 &                          &   distribution,      &  IMF distribution,   & high mass cores, & detections,\\
  &                          &   $\alpha_{\rm He}$      &  $\alpha_{\rm single\_IMF}$    & $R_{\rm massive} $ & $P_{\rm 0,massive} \times 10^{3} $ \\
\hline
 \\[0.0015ex]

\multicolumn{5}{l}{\it \bf Variations of assumptions affecting the entire population} 
 \\[0.01ex]
\multicolumn{5}{l}{\it  --  Exclusion of failed SNe} 
 \\[0.02ex]

(52) &  $M_{\mathrm{max,ccSN}} \sim 15 \Msun$                   &   -2.02 &  -2.45 &   0.29\tablefootmark{a} &  1000.0\tablefootmark{b} \\
(53) &  $M_{\mathrm{max,ccSN}} \sim 16.5 \Msun$                 &   -2.02 &  -2.42 &   0.2\tablefootmark{a}&  1000.0\tablefootmark{b} \\
(54) &  $M_{\mathrm{max,ccSN}} \sim 18 \Msun$                   &   -2.01 &  -2,40 &   0.18\tablefootmark{a} &   580.7\tablefootmark{b} \\
 29 &  $M_{\mathrm{max,ccSN}} \sim 20 \Msun$                   &   -1.97 &  -2.33 &   0.13\tablefootmark{a} &    67.8\tablefootmark{b} \\
\multicolumn{5}{l}{\it  -- Minimum mass for SNe} 
 \\[0.02ex]
	30 &  $M_{\mathrm{min,ccSN}} \sim 7 \Msun$                    &   -1.88 &  -2.20 &   0.19 &    7.9 \\
	31 &  $M_{\mathrm{min,ccSN}} \sim 8 \Msun$                    &   -1.89 &  -2.18 &   0.22 &    3.0 \\
\multicolumn{5}{l}{\it  --  IMF slope for massive stars} 
 \\[0.01ex]
  32 &  $\alpha_{\mathrm IMF}=-1.6$                             &   -1.36 &  -1.53 &   0.28 &     0.6 \\
  33 &  $\alpha_{\mathrm IMF}=-2.7$                             &   -2.21 &  -2.61 &   0.17 &     11.9 \\
  34 &  $\alpha_{\mathrm IMF}=-3.0$                             &   -2.44 &  -2.90 &   0.14 &    22.6 \\

\multicolumn{5}{l}{\it  --  Wind efficiency} 
 \\[0.01ex]
(51) &  wind factor $\eta = 0.1$                                &   -1.76 &  -1.98 &   0.32 &     0.3 \\
  25 &  wind factor $\eta = 0.33$                               &   -1.80 &  -2.04 &   0.30 &     0.4 \\
  (52) &  wind factor $\eta = 0.5$                               &   -1.80 &  -2.04 &   0.30 &     0.4 \\
  (53) &  wind factor $\eta = 1.5$                               &   -2.12 &  -2.50 &   0.12 &     48.7 \\
  (54) &  wind factor $\eta = 2.0$                               &   -3.56 &  -4.35 &   0.01 &     875.5 \\
  26 &  wind factor $\eta= 3.0$                                 &   -4.02 &  -4.95 &   0.00 &   879.1 \\
\multicolumn{5}{l}{\it   --  Metallicity} 
 \\[0.01ex]
  40 &  $Z=0.001$                                               &   -1.80 &  -2.07 &   0.29 &     0.6 \\
  42 &  $Z=0.008$                                               &   -1.78 &  -2.05 &   0.27 &     1.0 \\
  44 &  $Z=0.03$                                                &   -2.59 &  -3.10 &   0.03 &   404.2\\
 \hline
 \\
  \end{tabular}
\\
$({a})$ The mass threshold of 16.5 \Msun  to calculate $R_{\rm massive}$ is replaced by $M_{\mathrm{max,ccSN}}$.
$({b})$ The maximum equivalent single-star mass for Type II SNe to calculate this possibility is replaced by $M_{\mathrm{max,ccSN}}$, thus by definition increasing the possibility of zero detections of progenitors above 16.5 \Msun.
    }
\end{table*}

\begin{acknowledgements}

We are grateful to Rob Izzard for providing the population synthesis code, $\tt{binary\_c}$, used in this study and for his technical support. We are also grateful for the very useful discussions with Katie Auchettl, Ben Davies, Maria Drout, Tassos Fragos, Avishai Gal-Yam, Ylva G\"otberg, Avishai Gilkis, George Meynet, and Silvia Toonen. This project was funded in part by the European Union's Horizon 2020 research and innovation program from the European Research Council (ERC, Grant agreement No. 715063), and by the Netherlands Organization for Scientific Research (NWO) as part of the Vidi research program BinWaves with project number 639.042.728. EZ acknowledges support from the the Swiss National Science Foundation Professorship grant (project number PP00P2 176868) and from the Federal Commission for Scholarships for Foreign Students for the Swiss Government Excellence Scholarship (ESKAS No. 2019.0091). 

\end{acknowledgements}

\bibliography{my_bib,manos_bib_paper4}

\begin{thebibliography}{150}
\expandafter\ifx\csname natexlab\endcsname\relax\def\natexlab#1{#1}\fi

\bibitem[{{Adams} {et~al.}(2017{\natexlab{a}}){Adams}, {Kochanek}, {Gerke}, \&
  {Stanek}}]{Adams+2017a}
{Adams}, S.~M., {Kochanek}, C.~S., {Gerke}, J.~R., \& {Stanek}, K.~Z.
  2017{\natexlab{a}}, \mnras, 469, 1445

\bibitem[{{Adams} {et~al.}(2017{\natexlab{b}}){Adams}, {Kochanek}, {Gerke},
  {Stanek}, \& {Dai}}]{Adams+2017}
{Adams}, S.~M., {Kochanek}, C.~S., {Gerke}, J.~R., {Stanek}, K.~Z., \& {Dai},
  X. 2017{\natexlab{b}}, \mnras, 468, 4968

\bibitem[{{Aghakhanloo} {et~al.}(2017){Aghakhanloo}, {Murphy}, {Smith}, \&
  {Hlo{\v{z}}ek}}]{Aghakhanloo+2017}
{Aghakhanloo}, M., {Murphy}, J.~W., {Smith}, N., \& {Hlo{\v{z}}ek}, R. 2017,
  \mnras, 472, 591

\bibitem[{{Almeida} {et~al.}(2017){Almeida}, {Sana}, {Taylor}, {Barb{\'a}},
  {Bonanos}, {Crowther}, {Damineli}, {de Koter}, {de Mink}, {Evans}, {Gieles},
  {Grin}, {H{\'e}nault-Brunet}, {Langer}, {Lennon}, {Lockwood}, {Ma{\'{\i}}z
  Apell{\'a}niz}, {Moffat}, {Neijssel}, {Norman}, {Ram{\'{\i}}rez-Agudelo},
  {Richardson}, {Schootemeijer}, {Shenar}, {Soszy{\'n}ski}, {Tramper}, \&
  {Vink}}]{Almeida+2017}
{Almeida}, L.~A., {Sana}, H., {Taylor}, W., {et~al.} 2017, \aap, 598, A84

\bibitem[{{Anderson} {et~al.}(2018){Anderson}, {Dessart}, {Guti{\'e}rrez},
  {Kr{\"u}hler}, {Galbany}, {Jerkstrand}, {Smartt}, {Contreras}, {Morrell},
  {Phillips}, {Stritzinger}, {Hsiao}, {Gonz{\'a}lez-Gait{\'a}n}, {Agliozzo},
  {Castell{\'o}n}, {Chambers}, {Chen}, {Flewelling}, {Gonzalez},
  {Hosseinzadeh}, {Huber}, {Fraser}, {Inserra}, {Kankare}, {Mattila},
  {Magnier}, {Maguire}, {Lowe}, {Sollerman}, {Sullivan}, {Young}, \&
  {Valenti}}]{Anderson+2018}
{Anderson}, J.~P., {Dessart}, L., {Guti{\'e}rrez}, C.~P., {et~al.} 2018, Nature
  Astronomy, 2, 574

\bibitem[{{Anderson} {et~al.}(2012){Anderson}, {Habergham}, {James}, \&
  {Hamuy}}]{Anderson+2012}
{Anderson}, J.~P., {Habergham}, S.~M., {James}, P.~A., \& {Hamuy}, M. 2012,
  \mnras, 424, 1372

\bibitem[{{Asplund} {et~al.}(2009){Asplund}, {Grevesse}, {Sauval}, \&
  {Scott}}]{Asplund+2009}
{Asplund}, M., {Grevesse}, N., {Sauval}, A.~J., \& {Scott}, P. 2009, \araa, 47,
  481

\bibitem[{{Auchettl} {et~al.}(2019){Auchettl}, {Lopez}, {Badenes},
  {Ramirez-Ruiz}, {Beacom}, \& {Holland -Ashford}}]{Auchettl+2019}
{Auchettl}, K., {Lopez}, L.~A., {Badenes}, C., {et~al.} 2019, \apj, 871, 64

\bibitem[{{Baade} \& {Zwicky}(1934)}]{Baade+1934}
{Baade}, W. \& {Zwicky}, F. 1934, Proceedings of the National Academy of
  Science, 20, 254

\bibitem[{{Beasor} \& {Davies}(2016)}]{Beasor+2016}
{Beasor}, E.~R. \& {Davies}, B. 2016, \mnras, 463, 1269

\bibitem[{{Bethe} {et~al.}(1979){Bethe}, {Brown}, {Applegate}, \&
  {Lattimer}}]{Bethe+1979}
{Bethe}, H.~A., {Brown}, G.~E., {Applegate}, J., \& {Lattimer}, J.~M. 1979,
  Nuclear Physics A, 324, 487

\bibitem[{{Blaauw}(1961)}]{Blaauw1961}
{Blaauw}, A. 1961, \bain, 15, 265

\bibitem[{{Braun} \& {Langer}(1995)}]{Braun+1995}
{Braun}, H. \& {Langer}, N. 1995, \aap, 297, 483

\bibitem[{{Brott} {et~al.}(2011{\natexlab{a}}){Brott}, {de Mink}, {Cantiello},
  {Langer}, {de Koter}, {Evans}, {Hunter}, {Trundle}, \& {Vink}}]{Brott+2011}
{Brott}, I., {de Mink}, S.~E., {Cantiello}, M., {et~al.} 2011{\natexlab{a}},
  \aap, 530, A115

\bibitem[{{Brott} {et~al.}(2011{\natexlab{b}}){Brott}, {Evans}, {Hunter}, {de
  Koter}, {Langer}, {Dufton}, {Cantiello}, {Trundle}, {Lennon}, {de Mink},
  {Yoon}, \& {Anders}}]{Brott+2011a}
{Brott}, I., {Evans}, C.~J., {Hunter}, I., {et~al.} 2011{\natexlab{b}}, \aap,
  530, A116

\bibitem[{{Cao} {et~al.}(2013){Cao}, {Kasliwal}, {Arcavi}, {Horesh}, {Hancock},
  {Valenti}, {Cenko}, {Kulkarni}, {Gal-Yam}, {Gorbikov}, {Ofek}, {Sand},
  {Yaron}, {Graham}, {Silverman}, {Wheeler}, {Marion}, {Walker}, {Mazzali},
  {Howell}, {Li}, {Kong}, {Bloom}, {Nugent}, {Surace}, {Masci}, {Carpenter},
  {Degenaar}, \& {Gelino}}]{Cao+2013}
{Cao}, Y., {Kasliwal}, M.~M., {Arcavi}, I., {et~al.} 2013, \apjl, 775, L7

\bibitem[{{Chini} {et~al.}(2012){Chini}, {Hoffmeister}, {Nasseri}, {Stahl}, \&
  {Zinnecker}}]{Chini+2012}
{Chini}, R., {Hoffmeister}, V.~H., {Nasseri}, A., {Stahl}, O., \& {Zinnecker},
  H. 2012, \mnras, 424, 1925

\bibitem[{{Claeys} {et~al.}(2011){Claeys}, {de Mink}, {Pols}, {Eldridge}, \&
  {Baes}}]{Claeys+2011}
{Claeys}, J.~S.~W., {de Mink}, S.~E., {Pols}, O.~R., {Eldridge}, J.~J., \&
  {Baes}, M. 2011, \aap, 528, A131

\bibitem[{{Claret}(2007)}]{Claret2007a}
{Claret}, A. 2007, \aap, 475, 1019

\bibitem[{{Conti}(1975)}]{Conti1975}
{Conti}, P.~S. 1975, Memoires of the Societe Royale des Sciences de Liege, 9,
  193

\bibitem[{{Crawford}(1955)}]{Crawford1955}
{Crawford}, J.~A. 1955, \apj, 121, 71

\bibitem[{{Davies} \& {Beasor}(2018)}]{Davies+2018}
{Davies}, B. \& {Beasor}, E.~R. 2018, \mnras, 474, 2116

\bibitem[{{Davies} \& {Beasor}(2020)}]{Davies+2020}
{Davies}, B. \& {Beasor}, E.~R. 2020, \mnras, 170

\bibitem[{{De Donder} \& {Vanbeveren}(1998)}]{De-Donder+1998}
{De Donder}, E. \& {Vanbeveren}, D. 1998, \aap, 333, 557

\bibitem[{{De Donder} \& {Vanbeveren}(2003)}]{De-Donder+2003}
{De Donder}, E. \& {Vanbeveren}, D. 2003, \na, 8, 415

\bibitem[{{de Mink} {et~al.}(2013){de Mink}, {Langer}, {Izzard}, {Sana}, \& {de
  Koter}}]{de-Mink+2013}
{de Mink}, S.~E., {Langer}, N., {Izzard}, R.~G., {Sana}, H., \& {de Koter}, A.
  2013, \apj, 764, 166

\bibitem[{{Dewi} \& {Tauris}(2000)}]{Dewi+2000}
{Dewi}, J.~D.~M. \& {Tauris}, T.~M. 2000, \aap, 360, 1043

\bibitem[{{Dewi} \& {Tauris}(2001)}]{Dewi+2001}
{Dewi}, J.~D.~M. \& {Tauris}, T.~M. 2001, in Astronomical Society of the
  Pacific Conference Series, Vol. 229, Evolution of Binary and Multiple Star
  Systems, ed. P.~{Podsiadlowski}, S.~{Rappaport}, A.~R. {King}, F.~{D'Antona},
  \& L.~{Burderi}, 255

\bibitem[{{D{\'{\i}}az-Rodr{\'{\i}}guez}
  {et~al.}(2018){D{\'{\i}}az-Rodr{\'{\i}}guez}, {Murphy}, {Rubin}, {Dolphin},
  {Williams}, \& {Dalcanton}}]{Diaz-Rodriguez+2018}
{D{\'{\i}}az-Rodr{\'{\i}}guez}, M., {Murphy}, J.~W., {Rubin}, D.~A., {et~al.}
  2018, \apj, 861, 92

\bibitem[{{Dray} \& {Tout}(2007)}]{Dray+2007}
{Dray}, L.~M. \& {Tout}, C.~A. 2007, \mnras, 376, 61

\bibitem[{{Drout} {et~al.}(2011){Drout}, {Soderberg}, {Gal-Yam}, {Cenko},
  {Fox}, {Leonard}, {Sand}, {Moon}, {Arcavi}, \& {Green}}]{Drout+2011}
{Drout}, M.~R., {Soderberg}, A.~M., {Gal-Yam}, A., {et~al.} 2011, \apj, 741, 97

\bibitem[{{Duch{\^e}ne} \& {Kraus}(2013)}]{Duchene+2013}
{Duch{\^e}ne}, G. \& {Kraus}, A. 2013, \araa, 51, 269

\bibitem[{{Dunstall} {et~al.}(2015){Dunstall}, {Dufton}, {Sana}, {Evans},
  {Howarth}, {Sim{\'o}n-D{\'{\i}}az}, {de Mink}, {Langer}, {Ma{\'{\i}}z
  Apell{\'a}niz}, \& {Taylor}}]{Dunstall+2015}
{Dunstall}, P.~R., {Dufton}, P.~L., {Sana}, H., {et~al.} 2015, \aap, 580, A93

\bibitem[{{Eggleton} \& {Tokovinin}(2008)}]{Eggleton+2008}
{Eggleton}, P.~P. \& {Tokovinin}, A.~A. 2008, \mnras, 389, 869

\bibitem[{{Eldridge} {et~al.}(2013){Eldridge}, {Fraser}, {Smartt}, {Maund}, \&
  {Crockett}}]{Eldridge+2013}
{Eldridge}, J.~J., {Fraser}, M., {Smartt}, S.~J., {Maund}, J.~R., \&
  {Crockett}, R.~M. 2013, \mnras, 436, 774

\bibitem[{{Eldridge} {et~al.}(2019){Eldridge}, {Guo}, {Rodrigues}, {Stanway},
  \& {Xiao}}]{Eldridge+2019}
{Eldridge}, J.~J., {Guo}, N.~Y., {Rodrigues}, N., {Stanway}, E.~R., \& {Xiao},
  L. 2019, \pasa, 36, e041

\bibitem[{{Eldridge} {et~al.}(2011){Eldridge}, {Langer}, \&
  {Tout}}]{Eldridge+2011}
{Eldridge}, J.~J., {Langer}, N., \& {Tout}, C.~A. 2011, \mnras, 414, 3501

\bibitem[{{Eldridge} \& {Stanway}(2009)}]{Eldridge+2009}
{Eldridge}, J.~J. \& {Stanway}, E.~R. 2009, \mnras, 400, 1019

\bibitem[{{Eldridge} \& {Tout}(2004)}]{Eldridge+2004}
{Eldridge}, J.~J. \& {Tout}, C.~A. 2004, \mnras, 353, 87

\bibitem[{{Eldridge} {et~al.}(2018){Eldridge}, {Xiao}, {Stanway}, {Rodrigues},
  \& {Guo}}]{Eldridge+2018}
{Eldridge}, J.~J., {Xiao}, L., {Stanway}, E.~R., {Rodrigues}, N., \& {Guo},
  N.~Y. 2018, \pasa, 35, 49

\bibitem[{{Ertl} {et~al.}(2016){Ertl}, {Janka}, {Woosley}, {Sukhbold}, \&
  {Ugliano}}]{Ertl+2016}
{Ertl}, T., {Janka}, H.-T., {Woosley}, S.~E., {Sukhbold}, T., \& {Ugliano}, M.
  2016, \apj, 818, 124

\bibitem[{{Farrell} {et~al.}(2020){Farrell}, {Groh}, {Meynet}, \&
  {Eldridge}}]{Farrell+2020}
{Farrell}, E., {Groh}, J., {Meynet}, G., \& {Eldridge}, J. 2020, arXiv
  e-prints, arXiv:2001.08711

\bibitem[{{Filippenko}(1997)}]{Filippenko1997}
{Filippenko}, A.~V. 1997, \araa, 35, 309

\bibitem[{{Gal-Yam}(2017)}]{Gal-Yam2017}
{Gal-Yam}, A. 2017, {Observational and Physical Classification of Supernovae},
  ed. A.~W. {Alsabti} \& P.~{Murdin}, 195

\bibitem[{{Ge} {et~al.}(2010){Ge}, {Hjellming}, {Webbink}, {Chen}, \&
  {Han}}]{Ge+2010}
{Ge}, H., {Hjellming}, M.~S., {Webbink}, R.~F., {Chen}, X., \& {Han}, Z. 2010,
  \apj, 717, 724

\bibitem[{{Georgy} {et~al.}(2012){Georgy}, {Ekstr{\"o}m}, {Meynet}, {Massey},
  {Levesque}, {Hirschi}, {Eggenberger}, \& {Maeder}}]{Georgy+2012}
{Georgy}, C., {Ekstr{\"o}m}, S., {Meynet}, G., {et~al.} 2012, \aap, 542, A29

\bibitem[{{Gerke} {et~al.}(2015){Gerke}, {Kochanek}, \& {Stanek}}]{Gerke+2015}
{Gerke}, J.~R., {Kochanek}, C.~S., \& {Stanek}, K.~Z. 2015, \mnras, 450, 3289

\bibitem[{{Giannone}(1967)}]{Giannone1967}
{Giannone}, P. 1967, \zap, 65, 226

\bibitem[{{Graur} {et~al.}(2017){Graur}, {Bianco}, {Modjaz}, {Shivvers},
  {Filippenko}, {Li}, \& {Smith}}]{Graur+2017}
{Graur}, O., {Bianco}, F.~B., {Modjaz}, M., {et~al.} 2017, \apj, 837, 121

\bibitem[{{Habergham} {et~al.}(2014){Habergham}, {Anderson}, {James}, \&
  {Lyman}}]{Habergham+2014}
{Habergham}, S.~M., {Anderson}, J.~P., {James}, P.~A., \& {Lyman}, J.~D. 2014,
  \mnras, 441, 2230

\bibitem[{{Hamann} {et~al.}(1995){Hamann}, {Koesterke}, \&
  {Wessolowski}}]{Hamann+1995}
{Hamann}, W.-R., {Koesterke}, L., \& {Wessolowski}, U. 1995, \aap, 299, 151

\bibitem[{{Heger} {et~al.}(2003){Heger}, {Fryer}, {Woosley}, {Langer}, \&
  {Hartmann}}]{Heger+2003}
{Heger}, A., {Fryer}, C.~L., {Woosley}, S.~E., {Langer}, N., \& {Hartmann},
  D.~H. 2003, \apj, 591, 288

\bibitem[{{Hellings}(1983)}]{Hellings1983}
{Hellings}, P. 1983, \apss, 96, 37

\bibitem[{{Hellings}(1984)}]{Hellings1984}
{Hellings}, P. 1984, \apss, 104, 83

\bibitem[{{Hobbs} {et~al.}(2005){Hobbs}, {Lorimer}, {Lyne}, \&
  {Kramer}}]{Hobbs+2005}
{Hobbs}, G., {Lorimer}, D.~R., {Lyne}, A.~G., \& {Kramer}, M. 2005, \mnras,
  360, 974

\bibitem[{{Hopkins} {et~al.}(2014){Hopkins}, {Kere{\v s}}, {O{\~n}orbe},
  {Faucher-Gigu{\`e}re}, {Quataert}, {Murray}, \& {Bullock}}]{Hopkins+2014}
{Hopkins}, P.~F., {Kere{\v s}}, D., {O{\~n}orbe}, J., {et~al.} 2014, \mnras,
  445, 581

\bibitem[{{Humphreys} \& {Davidson}(1979)}]{Humphreys+1979}
{Humphreys}, R.~M. \& {Davidson}, K. 1979, \apj, 232, 409

\bibitem[{{Hurley} {et~al.}(2000){Hurley}, {Pols}, \& {Tout}}]{Hurley+2000}
{Hurley}, J.~R., {Pols}, O.~R., \& {Tout}, C.~A. 2000, \mnras, 315, 543

\bibitem[{{Hurley} {et~al.}(2002){Hurley}, {Tout}, \& {Pols}}]{Hurley+2002}
{Hurley}, J.~R., {Tout}, C.~A., \& {Pols}, O.~R. 2002, \mnras, 329, 897

\bibitem[{{Hut}(1980)}]{Hut1980}
{Hut}, P. 1980, \aap, 92, 167

\bibitem[{{Hut}(1981)}]{Hut1981}
{Hut}, P. 1981, \aap, 99, 126

\bibitem[{{Izzard} {et~al.}(2006){Izzard}, {Dray}, {Karakas}, {Lugaro}, \&
  {Tout}}]{Izzard+2006}
{Izzard}, R.~G., {Dray}, L.~M., {Karakas}, A.~I., {Lugaro}, M., \& {Tout},
  C.~A. 2006, \aap, 460, 565

\bibitem[{{Izzard} {et~al.}(2009){Izzard}, {Glebbeek}, {Stancliffe}, \&
  {Pols}}]{Izzard+2009}
{Izzard}, R.~G., {Glebbeek}, E., {Stancliffe}, R.~J., \& {Pols}, O.~R. 2009,
  \aap, 508, 1359

\bibitem[{{Izzard} {et~al.}(2004){Izzard}, {Tout}, {Karakas}, \&
  {Pols}}]{Izzard+2004}
{Izzard}, R.~G., {Tout}, C.~A., {Karakas}, A.~I., \& {Pols}, O.~R. 2004,
  \mnras, 350, 407

\bibitem[{{Janka}(2012)}]{Janka2012}
{Janka}, H.-T. 2012, Annual Review of Nuclear and Particle Science, 62, 407

\bibitem[{{Jennings} {et~al.}(2014){Jennings}, {Williams}, {Murphy},
  {Dalcanton}, {Gilbert}, {Dolphin}, {Weisz}, \& {Fouesneau}}]{Jennings+2014}
{Jennings}, Z.~G., {Williams}, B.~F., {Murphy}, J.~W., {et~al.} 2014, \apj,
  795, 170

\bibitem[{{Jerkstrand} {et~al.}(2012){Jerkstrand}, {Fransson}, {Maguire},
  {Smartt}, {Ergon}, \& {Spyromilio}}]{Jerkstrand+2012}
{Jerkstrand}, A., {Fransson}, C., {Maguire}, K., {et~al.} 2012, \aap, 546, A28

\bibitem[{{Jerkstrand} {et~al.}(2014){Jerkstrand}, {Smartt}, {Fraser},
  {Fransson}, {Sollerman}, {Taddia}, \& {Kotak}}]{Jerkstrand+2014}
{Jerkstrand}, A., {Smartt}, S.~J., {Fraser}, M., {et~al.} 2014, \mnras, 439,
  3694

\bibitem[{{Justham} {et~al.}(2014){Justham}, {Podsiadlowski}, \&
  {Vink}}]{Justham+2014}
{Justham}, S., {Podsiadlowski}, P., \& {Vink}, J.~S. 2014, \apj, 796, 121

\bibitem[{{Kilpatrick} \& {Foley}(2018)}]{Kilpatrick+2018}
{Kilpatrick}, C.~D. \& {Foley}, R.~J. 2018, \mnras, 481, 2536

\bibitem[{{Kiminki} \& {Kobulnicky}(2012)}]{Kiminki+2012}
{Kiminki}, D.~C. \& {Kobulnicky}, H.~A. 2012, \apj, 751, 4

\bibitem[{{Kobulnicky} \& {Fryer}(2007)}]{Kobulnicky+2007}
{Kobulnicky}, H.~A. \& {Fryer}, C.~L. 2007, \apj, 670, 747

\bibitem[{{Kochanek} {et~al.}(2008){Kochanek}, {Beacom}, {Kistler}, {Prieto},
  {Stanek}, {Thompson}, \& {Y{\"u}ksel}}]{Kochanek+2008}
{Kochanek}, C.~S., {Beacom}, J.~F., {Kistler}, M.~D., {et~al.} 2008, \apj, 684,
  1336

\bibitem[{{Kochanek} {et~al.}(2012){Kochanek}, {Khan}, \&
  {Dai}}]{Kochanek+2012}
{Kochanek}, C.~S., {Khan}, R., \& {Dai}, X. 2012, \apj, 759, 20

\bibitem[{{Kroupa}(2001)}]{Kroupa2001}
{Kroupa}, P. 2001, \mnras, 322, 231

\bibitem[{{Kuncarayakti} {et~al.}(2018){Kuncarayakti}, {Anderson}, {Galbany},
  {Maeda}, {Hamuy}, {Aldering}, {Arimoto}, {Doi}, {Morokuma}, \&
  {Usuda}}]{Kuncarayakti+2018}
{Kuncarayakti}, H., {Anderson}, J.~P., {Galbany}, L., {et~al.} 2018, \aap, 613,
  A35

\bibitem[{{Laplace} {et~al.}(2020){Laplace}, {G{\"o}tberg}, {de Mink},
  {Justham}, \& {Farmer}}]{Laplace+2020}
{Laplace}, E., {G{\"o}tberg}, Y., {de Mink}, S.~E., {Justham}, S., \& {Farmer},
  R. 2020, \aap, 637, A6

\bibitem[{{Larson}(1974)}]{Larson1974}
{Larson}, R.~B. 1974, \mnras, 169, 229

\bibitem[{{Li} {et~al.}(2011){Li}, {Leaman}, {Chornock}, {Filippenko},
  {Poznanski}, {Ganeshalingam}, {Wang}, {Modjaz}, {Jha}, {Foley}, \&
  {Smith}}]{Li+2011}
{Li}, W., {Leaman}, J., {Chornock}, R., {et~al.} 2011, \mnras, 412, 1441

\bibitem[{{Lyman} {et~al.}(2016){Lyman}, {Bersier}, {James}, {Mazzali},
  {Eldridge}, {Fraser}, \& {Pian}}]{Lyman+2016}
{Lyman}, J.~D., {Bersier}, D., {James}, P.~A., {et~al.} 2016, \mnras, 457, 328

\bibitem[{{Maoz} \& {Badenes}(2010)}]{Maoz+2010}
{Maoz}, D. \& {Badenes}, C. 2010, \mnras, 407, 1314

\bibitem[{{Martinez} \& {Bersten}(2019)}]{Martinez+2019}
{Martinez}, L. \& {Bersten}, M.~C. 2019, \aap, 629, A124

\bibitem[{{Maund}(2017)}]{Maund2017}
{Maund}, J.~R. 2017, \mnras, 469, 2202

\bibitem[{{Maund} \& {Smartt}(2005)}]{Maund+2005a}
{Maund}, J.~R. \& {Smartt}, S.~J. 2005, \mnras, 360, 288

\bibitem[{{Menon} \& {Heger}(2017)}]{Menon+2017}
{Menon}, A. \& {Heger}, A. 2017, \mnras, 469, 4649

\bibitem[{{Menon} {et~al.}(2019){Menon}, {Utrobin}, \& {Heger}}]{Menon+2019}
{Menon}, A., {Utrobin}, V., \& {Heger}, A. 2019, \mnras, 482, 438

\bibitem[{{Mestel}(1957)}]{Mestel1957}
{Mestel}, L. 1957, \apj, 126, 550

\bibitem[{{Mestel} \& {Moss}(1986)}]{Mestel+1986}
{Mestel}, L. \& {Moss}, D.~L. 1986, \mnras, 221, 25

\bibitem[{{Moe} \& {Di Stefano}(2017)}]{Moe+2017}
{Moe}, M. \& {Di Stefano}, R. 2017, \apjs, 230, 15

\bibitem[{{Mokiem} {et~al.}(2007){Mokiem}, {de Koter}, {Evans}, {Puls},
  {Smartt}, {Crowther}, {Herrero}, {Langer}, {Lennon}, {Najarro}, {Villamariz},
  \& {Vink}}]{Mokiem+2007}
{Mokiem}, M.~R., {de Koter}, A., {Evans}, C.~J., {et~al.} 2007, \aap, 465, 1003

\bibitem[{{Murphy} {et~al.}(2011){Murphy}, {Jennings}, {Williams}, {Dalcanton},
  \& {Dolphin}}]{Murphy+2011}
{Murphy}, J.~W., {Jennings}, Z.~G., {Williams}, B., {Dalcanton}, J.~J., \&
  {Dolphin}, A.~E. 2011, \apjl, 742, L4

\bibitem[{{Nieuwenhuijzen} \& {de Jager}(1990)}]{Nieuwenhuijzen+1990}
{Nieuwenhuijzen}, H. \& {de Jager}, C. 1990, \aap, 231, 134

\bibitem[{{O'Connor} \& {Ott}(2011)}]{OConnor+2011}
{O'Connor}, E. \& {Ott}, C.~D. 2011, \apj, 730, 70

\bibitem[{{{\"O}pik}(1924)}]{Opik1924}
{{\"O}pik}, E. 1924, Tartu Obs. Publ., 25, 6

\bibitem[{{Pavlovskii} \& {Ivanova}(2015)}]{Pavlovskii+2015}
{Pavlovskii}, K. \& {Ivanova}, N. 2015, \mnras, 449, 4415

\bibitem[{{Pavlovskii} {et~al.}(2017){Pavlovskii}, {Ivanova}, {Belczynski}, \&
  {Van}}]{Pavlovskii+2017}
{Pavlovskii}, K., {Ivanova}, N., {Belczynski}, K., \& {Van}, K.~X. 2017,
  \mnras, 465, 2092

\bibitem[{{Piro}(2013)}]{Piro2013}
{Piro}, A.~L. 2013, \apjl, 768, L14

\bibitem[{{Podsiadlowski}(1992)}]{Podsiadlowski1992}
{Podsiadlowski}, P. 1992, \pasp, 104, 717

\bibitem[{{Podsiadlowski} {et~al.}(1992){Podsiadlowski}, {Joss}, \&
  {Hsu}}]{Podsiadlowski+1992}
{Podsiadlowski}, P., {Joss}, P.~C., \& {Hsu}, J.~J.~L. 1992, \apj, 391, 246

\bibitem[{{Podsiadlowski} {et~al.}(1990){Podsiadlowski}, {Joss}, \&
  {Rappaport}}]{Podsiadlowski+1990}
{Podsiadlowski}, P., {Joss}, P.~C., \& {Rappaport}, S. 1990, \aap, 227, L9

\bibitem[{{Pols} {et~al.}(1998){Pols}, {Schr\"oder}, {Hurley}, {Tout}, \&
  {Eggleton}}]{Pols+1998}
{Pols}, O.~R., {Schr\"oder}, K.-P., {Hurley}, J.~R., {Tout}, C.~A., \&
  {Eggleton}, P.~P. 1998, \mnras, 298, 525

\bibitem[{{Prentice} {et~al.}(2019){Prentice}, {Ashall}, {James}, {Short},
  {Mazzali}, {Bersier}, {Crowther}, {Barbarino}, {Chen}, {Copperwheat},
  {Darnley}, {Denneau}, {Elias-Rosa}, {Fraser}, {Galbany}, {Gal-Yam},
  {Harmanen}, {Howell}, {Hosseinzadeh}, {Inserra}, {Kankare}, {Karamehmetoglu},
  {Lamb}, {Limongi}, {Maguire}, {McCully}, {Olivares E}, {Piascik}, {Pignata},
  {Reichart}, {Rest}, {Reynolds}, {Rodr{\'\i}guez}, {Saario}, {Schulze},
  {Smartt}, {Smith}, {Sollerman}, {Stalder}, {Sullivan}, {Taddia}, {Valenti},
  {Vergani}, {Williams}, \& {Young}}]{Prentice+2019}
{Prentice}, S.~J., {Ashall}, C., {James}, P.~A., {et~al.} 2019, \mnras, 485,
  1559

\bibitem[{{Renzo} {et~al.}(2017){Renzo}, {Ott}, {Shore}, \& {de
  Mink}}]{Renzo+2017}
{Renzo}, M., {Ott}, C.~D., {Shore}, S.~N., \& {de Mink}, S.~E. 2017, \aap, 603,
  A118

\bibitem[{{Renzo} {et~al.}(2019){Renzo}, {Zapartas}, {de Mink}, {G{\"o}tberg},
  {Justham}, {Farmer}, {Izzard}, {Toonen}, \& {Sana}}]{Renzo+2019}
{Renzo}, M., {Zapartas}, E., {de Mink}, S.~E., {et~al.} 2019, \aap, 624, A66

\bibitem[{{Ribas} {et~al.}(2000){Ribas}, {Jordi}, \&
  {Gim{\'e}nez}}]{Ribas+2000}
{Ribas}, I., {Jordi}, C., \& {Gim{\'e}nez}, {\'A}. 2000, \mnras, 318, L55

\bibitem[{{Rui} {et~al.}(2019){Rui}, {Wang}, {Mo}, {Xiang}, {Zhang}, {Maund},
  {Gal-Yam}, {Wang}, \& {Zhang}}]{Rui+2019}
{Rui}, L., {Wang}, X., {Mo}, J., {et~al.} 2019, \mnras, 485, 1990

\bibitem[{{Salpeter}(1955)}]{Salpeter1955}
{Salpeter}, E.~E. 1955, \apj, 121, 161

\bibitem[{Sana {et~al.}(2012)Sana, de~Mink, de~Koter, Langer, Evans, Gieles,
  Gosset, Izzard, Le~Bouquin, \& Schneider}]{Sana+2012}
Sana, H., de~Mink, S.~E., de~Koter, A., {et~al.} 2012, Science, 337, 444

\bibitem[{{Schady} {et~al.}(2019){Schady}, {Eldridge}, {Anderson}, {Chen},
  {Galbany}, {Kuncarayakti}, \& {Xiao}}]{Schady+2019}
{Schady}, P., {Eldridge}, J.~J., {Anderson}, J., {et~al.} 2019, \mnras, 490,
  4515

\bibitem[{{Schneider} {et~al.}(2015){Schneider}, {Izzard}, {Langer}, \& {de
  Mink}}]{Schneider+2015}
{Schneider}, F.~R.~N., {Izzard}, R.~G., {Langer}, N., \& {de Mink}, S.~E. 2015,
  \apj, 805, 20

\bibitem[{{Schneider} {et~al.}(2019){Schneider}, {Ohlmann}, {Podsiadlowski},
  {R{\"o}pke}, {Balbus}, {Pakmor}, \& {Springel}}]{Schneider+2019}
{Schneider}, F. R.~N., {Ohlmann}, S.~T., {Podsiadlowski}, P., {et~al.} 2019,
  \nat, 574, 211

\bibitem[{{Shao} \& {Li}(2014)}]{Shao+2014}
{Shao}, Y. \& {Li}, X.-D. 2014, \apj, 796, 37

\bibitem[{{Silverman} {et~al.}(2017){Silverman}, {Pickett}, {Wheeler},
  {Filippenko}, {Vink{\'o}}, {Marion}, {Cenko}, {Chornock}, {Clubb}, {Foley},
  {Graham}, {Kelly}, {Matheson}, \& {Shields}}]{Silverman+2017}
{Silverman}, J.~M., {Pickett}, S., {Wheeler}, J.~C., {et~al.} 2017, \mnras,
  467, 369

\bibitem[{{Smartt}(2015)}]{Smartt2015}
{Smartt}, S.~J. 2015, \pasa, 32, e016

\bibitem[{{Smartt} {et~al.}(2009){Smartt}, {Eldridge}, {Crockett}, \&
  {Maund}}]{Smartt+2009}
{Smartt}, S.~J., {Eldridge}, J.~J., {Crockett}, R.~M., \& {Maund}, J.~R. 2009,
  \mnras, 395, 1409

\bibitem[{{Smartt} {et~al.}(2001){Smartt}, {Gilmore}, {Trentham}, {Tout}, \&
  {Frayn}}]{Smartt+2001}
{Smartt}, S.~J., {Gilmore}, G.~F., {Trentham}, N., {Tout}, C.~A., \& {Frayn},
  C.~M. 2001, \apjl, 556, L29

\bibitem[{{Smith}(2014)}]{Smith2014}
{Smith}, N. 2014, \araa, 52, 487

\bibitem[{{Smith} {et~al.}(2011){Smith}, {Li}, {Filippenko}, \&
  {Chornock}}]{Smith+2011}
{Smith}, N., {Li}, W., {Filippenko}, A.~V., \& {Chornock}, R. 2011, \mnras,
  412, 1522

\bibitem[{{Smith} \& {Tombleson}(2015)}]{Smith+2015}
{Smith}, N. \& {Tombleson}, R. 2015, \mnras, 447, 598

\bibitem[{{Sravan} {et~al.}(2018){Sravan}, {Marchant}, {Kalogera}, \&
  {Margutti}}]{Sravan+2018}
{Sravan}, N., {Marchant}, P., {Kalogera}, V., \& {Margutti}, R. 2018, \apjl,
  852, L17

\bibitem[{{Straniero} {et~al.}(2019){Straniero}, {Dominguez}, {Piersanti},
  {Giannotti}, \& {Mirizzi}}]{Straniero+2019}
{Straniero}, O., {Dominguez}, I., {Piersanti}, L., {Giannotti}, M., \&
  {Mirizzi}, A. 2019, \apj, 881, 158

\bibitem[{{Sukhbold} {et~al.}(2016){Sukhbold}, {Ertl}, {Woosley}, {Brown}, \&
  {Janka}}]{Sukhbold+2016}
{Sukhbold}, T., {Ertl}, T., {Woosley}, S.~E., {Brown}, J.~M., \& {Janka}, H.-T.
  2016, \apj, 821, 38

\bibitem[{{Taddia} {et~al.}(2015){Taddia}, {Sollerman}, {Leloudas},
  {Stritzinger}, {Valenti}, {Galbany}, {Kessler}, {Schneider}, \&
  {Wheeler}}]{Taddia+2015}
{Taddia}, F., {Sollerman}, J., {Leloudas}, G., {et~al.} 2015, \aap, 574, A60

\bibitem[{{Tauris} \& {Dewi}(2001)}]{Tauris+2001}
{Tauris}, T.~M. \& {Dewi}, J.~D.~M. 2001, \aap, 369, 170

\bibitem[{{Tomasella} {et~al.}(2013){Tomasella}, {Cappellaro}, {Fraser},
  {Pumo}, {Pastorello}, {Pignata}, {Benetti}, {Bufano}, {Dennefeld},
  {Harutyunyan}, {Iijima}, {Jerkstrand }, {Kankare}, {Kotak}, {Magill},
  {Nascimbeni}, {Ochner}, {Siviero}, {Smartt}, {Sollerman}, {Stanishev},
  {Taddia}, {Taubenberger}, {Turatto}, {Valenti}, {Wright}, \&
  {Zampieri}}]{Tomasella+2013}
{Tomasella}, L., {Cappellaro}, E., {Fraser}, M., {et~al.} 2013, \mnras, 434,
  1636

\bibitem[{{Tout} {et~al.}(1997){Tout}, {Aarseth}, {Pols}, \&
  {Eggleton}}]{Tout+1997}
{Tout}, C.~A., {Aarseth}, S.~J., {Pols}, O.~R., \& {Eggleton}, P.~P. 1997,
  \mnras, 291, 732

\bibitem[{{Tutukov} {et~al.}(1992){Tutukov}, {Yungelson}, \&
  {Iben}}]{Tutukov+1992}
{Tutukov}, A.~V., {Yungelson}, L.~R., \& {Iben}, Icko, J. 1992, \apj, 386, 197

\bibitem[{{Valenti} {et~al.}(2016){Valenti}, {Howell}, {Stritzinger}, {Graham},
  {Hosseinzadeh}, {Arcavi}, {Bildsten}, {Jerkstrand}, {McCully}, {Pastorello},
  {Piro}, {Sand}, {Smartt}, {Terreran}, {Baltay}, {Benetti}, {Brown},
  {Filippenko}, {Fraser}, {Rabinowitz}, {Sullivan}, \& {Yuan}}]{Valenti+2016}
{Valenti}, S., {Howell}, D.~A., {Stritzinger}, M.~D., {et~al.} 2016, \mnras,
  459, 3939

\bibitem[{{Van Dyk} {et~al.}(2003){Van Dyk}, {Li}, \&
  {Filippenko}}]{Van-Dyk+2003}
{Van Dyk}, S.~D., {Li}, W., \& {Filippenko}, A.~V. 2003, \pasp, 115, 1

\bibitem[{{Van Dyk} {et~al.}(1999){Van Dyk}, {Peng}, {Barth}, \&
  {Filippenko}}]{Van-Dyk+1999}
{Van Dyk}, S.~D., {Peng}, C.~Y., {Barth}, A.~J., \& {Filippenko}, A.~V. 1999,
  \aj, 118, 2331

\bibitem[{{Van Dyk} {et~al.}(2018){Van Dyk}, {Zheng}, {Brink}, {Filippenko},
  {Milisavljevic}, {Andrews}, {Smith}, {Cignoni}, {Fox}, {Kelly}, {Adamo},
  {Yunus}, {Zhang}, \& {Kumar}}]{Van-Dyk+2018}
{Van Dyk}, S.~D., {Zheng}, W., {Brink}, T.~G., {et~al.} 2018, \apj, 860, 90

\bibitem[{{Van Dyk} {et~al.}(2019){Van Dyk}, {Zheng}, {Maund}, {Brink},
  {Srinivasan}, {Andrews}, {Smith}, {Leonard}, {Morozova}, {Filippenko},
  {Conner}, {Milisavljevic}, {de Jaeger}, {Long}, {Isaacson}, {Crossfield},
  {Kosiarek}, {Howard}, {Fox}, {Kelly}, {Piro}, {Littlefair}, {Dhillon},
  {Wilson}, {Butterley}, {Yunus}, {Channa}, {Jeffers}, {Falcon}, {Ross},
  {Hestenes}, {Stegman}, {Zhang}, \& {Kumar}}]{Van-Dyk+2019}
{Van Dyk}, S.~D., {Zheng}, W., {Maund}, J.~R., {et~al.} 2019, \apj, 875, 136

\bibitem[{{Vanbeveren} {et~al.}(2013){Vanbeveren}, {Mennekens}, {Van
  Rensbergen}, \& {De Loore}}]{Vanbeveren+2013}
{Vanbeveren}, D., {Mennekens}, N., {Van Rensbergen}, W., \& {De Loore}, C.
  2013, \aap, 552, A105

\bibitem[{{Verbunt} \& {Cator}(2017)}]{Verbunt+2017}
{Verbunt}, F. \& {Cator}, E. 2017, Journal of Astrophysics and Astronomy, 38,
  40

\bibitem[{{Vink} {et~al.}(2000){Vink}, {de Koter}, \& {Lamers}}]{Vink+2000}
{Vink}, J.~S., {de Koter}, A., \& {Lamers}, H.~J.~G.~L.~M. 2000, \aap, 362, 295

\bibitem[{{Vink} {et~al.}(2001){Vink}, {de Koter}, \& {Lamers}}]{Vink+2001}
{Vink}, J.~S., {de Koter}, A., \& {Lamers}, H.~J.~G.~L.~M. 2001, \aap, 369, 574

\bibitem[{{Walmswell} \& {Eldridge}(2012)}]{Walmswell+2012}
{Walmswell}, J.~J. \& {Eldridge}, J.~J. 2012, \mnras, 419, 2054

\bibitem[{{Webbink}(1984)}]{Webbink1984}
{Webbink}, R.~F. 1984, \apj, 277, 355

\bibitem[{{Williams} {et~al.}(2018){Williams}, {Hillis}, {Murphy}, {Gilbert},
  {Dalcanton}, \& {Dolphin}}]{Williams+2018}
{Williams}, B.~F., {Hillis}, T.~J., {Murphy}, J.~W., {et~al.} 2018, \apj, 860,
  39

\bibitem[{{Williams} {et~al.}(2014){Williams}, {Peterson}, {Murphy}, {Gilbert},
  {Dalcanton}, {Dolphin}, \& {Jennings}}]{Williams+2014}
{Williams}, B.~F., {Peterson}, S., {Murphy}, J., {et~al.} 2014, \apj, 791, 105

\bibitem[{{Woosley} {et~al.}(2002){Woosley}, {Heger}, \&
  {Weaver}}]{Woosley+2002}
{Woosley}, S.~E., {Heger}, A., \& {Weaver}, T.~A. 2002, Reviews of Modern
  Physics, 74, 1015

\bibitem[{{Xiao} {et~al.}(2019){Xiao}, {Galbany}, {Eldridge}, \&
  {Stanway}}]{Xiao+2019}
{Xiao}, L., {Galbany}, L., {Eldridge}, J.~J., \& {Stanway}, E.~R. 2019, \mnras,
  482, 384

\bibitem[{{Xiao} {et~al.}(2018){Xiao}, {Stanway}, \& {Eldridge}}]{Xiao+2018}
{Xiao}, L., {Stanway}, E.~R., \& {Eldridge}, J.~J. 2018, \mnras, 477, 904

\bibitem[{{Yoon} {et~al.}(2017){Yoon}, {Dessart}, \& {Clocchiatti}}]{Yoon+2017}
{Yoon}, S.-C., {Dessart}, L., \& {Clocchiatti}, A. 2017, \apj, 840, 10

\bibitem[{{Yoon} \& {Langer}(2005)}]{Yoon+2005}
{Yoon}, S.-C. \& {Langer}, N. 2005, \aap, 443, 643

\bibitem[{{Yoon} {et~al.}(2010){Yoon}, {Woosley}, \& {Langer}}]{Yoon+2010}
{Yoon}, S.-C., {Woosley}, S.~E., \& {Langer}, N. 2010, \apj, 725, 940

\bibitem[{{Zahn}(1977)}]{Zahn1977}
{Zahn}, J.-P. 1977, \aap, 57, 383

\bibitem[{{Zapartas} {et~al.}(2017{\natexlab{a}}){Zapartas}, {de Mink},
  {Izzard}, {Yoon}, {Badenes}, {G{\"o}tberg}, {de Koter}, {Neijssel}, {Renzo},
  {Schootemeijer}, \& {Shrotriya}}]{Zapartas+2017}
{Zapartas}, E., {de Mink}, S.~E., {Izzard}, R.~G., {et~al.} 2017{\natexlab{a}},
  \aap, 601, A29

\bibitem[{{Zapartas} {et~al.}(2019){Zapartas}, {de Mink}, {Justham}, {Smith},
  {de Koter}, {Renzo}, {Arcavi}, {Farmer}, {G{\"o}tberg}, \&
  {Toonen}}]{Zapartas+2019}
{Zapartas}, E., {de Mink}, S.~E., {Justham}, S., {et~al.} 2019, \aap, 631, A5

\bibitem[{{Zapartas} {et~al.}(2017{\natexlab{b}}){Zapartas}, {de Mink}, {Van
  Dyk}, {Fox}, {Smith}, {Bostroem}, {de Koter}, {Filippenko}, {Izzard},
  {Kelly}, {Neijssel}, {Renzo}, \& {Ryder}}]{Zapartas+2017a}
{Zapartas}, E., {de Mink}, S.~E., {Van Dyk}, S.~D., {et~al.}
  2017{\natexlab{b}}, \apj, 842, 125

\end{thebibliography}

\end{document}